\documentclass[12pt,a4j]{article}
\usepackage[dvips]{graphicx}
\usepackage{enumerate}
\usepackage{amsmath}
\usepackage{amsfonts}
\usepackage{mathrsfs}

\begin{document}
\baselineskip=5.5mm
\centerline{\bf Multiphase solutions and their reductions for a nonlocal  }\par
\centerline{\bf  nonlinear Schr\"odinger equation  with focusing nonlinearity}\par
\bigskip
\centerline{Yoshimasa Matsuno\footnote{{\it E-mail address}: matsuno@yamaguchi-u.ac.jp}}\par

\centerline{\it Division of Applied Mathematical Science,}\par
\centerline{\it Graduate School of Science and Technology for Innovation} \par
\centerline{\it Yamaguchi University, Ube, Yamaguchi 755-8611, Japan} \par
\bigskip
\bigskip
\leftline{\bf Abstract}\par
\noindent  A nonlocal nonlinear Schr\"odinger equation with focusing nonlinearity is considered which has been 
derived as a continuum limit of the 
Calogero-Sutherland model in an integrable classical dynamical system. The equation is shown to stem from the compatibility conditions of 
a system of linear PDEs, assuring its complete integrability.
 We construct a nonsingular $N$-phase solution ($N$: positive integer) of the equation by means of a direct
method.  The features of  the one- and two-phase  solutions are  investigated
in comparison with the corresponding solutions of the defocusing version of the equation.
We also provide an alternative representation of the $N$-phase solution
in terms of solutions of a system of nonlinear algebraic equations. Furthermore, the eigenvalue problem associated with the $N$-phase solution
is discussed briefly with some exact results. Subsequently, we demonstrate that the $N$-soliton solution can be obtained simply by taking the long-wave limit of the
$N$-phase solution. The similar limiting procedure gives an alternative representation of the $N$-soliton solution as well as the  exact results related to 
the corresponding  eigenvalue problem. 
\par
\newpage
\leftline{\bf  1. Introduction} \par
\bigskip
\noindent 1.1. Nonlocal nonlinear Schr\"odinger equation \par
\medskip
\noindent The  nonlocal nonlinear Schr\"odinger (NLS) equation that we consider here can be written in the form
$${\rm i}u_t=u_{xx}- {\rm i}u(1-{\rm i}H)(|u|^2)_x, \eqno(1.1a)$$
 with the operator $H$ being the Hilbert transform defined by
$$Hu(x, t)={1\over \pi}P\int_{-\infty}^\infty{u(y, t) \over y-x}\,dy, \quad x\in (-\infty, \infty), \eqno(1.1b)$$
for functions vanishing at infinity, and
$$Hu(x, t)={1\over 2L}P\int_{-L}^L{\cot}\left[{\pi(y-x)\over 2L}\right]u(y, t)\,dy, \quad x\in (-L, L), \eqno(1.1c)$$
for functions with period $2L$. Here, $u=u(x, t)$ is a complex function and the subscripts $t$ and $x$ appended to $u$ denote
partial differentiation.
Using the formulas $He^{{\rm i}kx}={\rm i}\ {\rm sgn}(k)\,e^{{\rm i}kx}\ (k\in \mathbb{R}),\ \sum_{n=1}^\infty\sin\, nx={1\over 2}\,\cot\,{x\over 2}$, where $H$ is defined by (1.1b),
one can show that both definitions of the Hilbert transform are identical when applied to the $2L$-periodic functions.
\par
 Equation (1.1) is characterized by the nonlocal nonlinearity of focusing type. It has been derived as a continuum limit of the 
Calogero-Sutherland model in an integrable classical dynamical system [1]. We recall that the defocusing version of equation (1.1)
$${\rm i}u_t=u_{xx}+ {\rm i}u(1-{\rm i}H)(|u|^2)_x, \eqno(1.2)$$
has been introduced as a deep-water limit of the intermediate NLS equation of defocusing type which describes the long-term evolution of 
the envelope of quasi-harmonic internal waves in a stratified fluid of finite depth [2].
A number of results have been obtained for equation (1.2).  Among them are multiperiodic and multisoliton solutions [3], linear stability of the multisoliton 
solutions [4], integrable dynamical systems associated with it [5],
asymptotic analysis of periodic solutions in the limit of small dispersion [6],  initial value
problems [7, 8] and new representation of multiperiodic solutions [9]. \par
  On the other hand, only a few outcomes are available for equation (1.1). To be more specific,
the one-soliton solution of equation (1.1)  has been obtained by performing a  limiting procedure to the
one-soliton solution of the intermediate NLS equation of focusing type [10]. The construction of the general $N$-soliton solution ($N$: positive integer), however,  remained open
until very recently.
In fact, a recent work  deals with the Lax integrability and 
multisoliton solutions as well as their dynamics [11]. As in the case of the NLS equation [12-14], the feature of solutions depends crucially on the types of nonlinearity.
In particular, the global-in-time existence and the growth of the Sobolev norms  have been demonstrated  for the  multisoliton soliton solutions of (1.1).
Whether the similar results can be established for the multiphase solutions is an interesting issue to be resolved in a future work.
Quite recently, the intermediate version of the focusing-defocusing Manakov system was proposed with some numerical solutions [15].
The focusing nonlocal NLS equation is now an interesting issue in search of integrable nonlocal nonlinear partial differential equations (PDEs). \par
\bigskip
\noindent 1.2. Integrability  \par
\medskip
\noindent Integrability of given nonlinear PDE may be characterized by the existence of the Lax pair
and infinite number of conservation lows. Actually, 
equation (1.1) was shown to exhibit a Lax pair structure and associated conservation laws [11].
One can also derive equation (1.1) as the compatibility conditions of the following system of linear PDEs for the 
eigenfunctions $\phi$ and $\psi^\pm$
$${\rm i}\phi_x+\lambda \phi+u\psi^+=0, \eqno(1.3)$$
$$\psi^+-\sigma \psi^--u^*\phi=0, \eqno(1.4)$$
$${\rm i}\phi_t-2{\rm i}\lambda \phi_x+\phi_{xx}-2{\rm i}u_x\psi^+-\kappa\phi=0, \eqno(1.5)$$
$${\rm i}\psi^\pm_t-2{\rm i}\lambda \psi^\pm_x+\psi^\pm_{xx}-{\rm i}[(\pm1-{\rm i}H)(|u|^2)_x-{\rm i}\kappa]\psi^\pm=0, \eqno(1.6)$$
where $\psi^+(\psi^-)$ is the boundary value of function analytic in the upper (lower)-half complex $x$ plane ${\rm Im\,} x>0\ ({\rm Im}\, x<0)$
and they have the unique representations $\psi^\pm=\pm {1\over 2}(1\mp{\rm i} H)\psi$ with $\psi$ being a complex function.
Here, the asterisk appended to $u$ denotes the complex conjugation, $\lambda$  is the spectral parameter
and $\sigma$ and $\kappa$ are constants related to $\lambda$ which are  fixed by the boundary conditions for $\phi$ and $\psi^\pm$.
Notice that the corresponding  linear system for equation (1.2) can be obtained from (1.3)-(1.6) if one replaces the minus sign of the third term on the left hand-side of
$(1.4)$ with the plus one. See, for example [7]. One can see that the linear system (1.3)-(1.6) yields a Lax pair equivalent to that obtained in [11].
\par
\bigskip
\noindent 1.3. Outline of the paper\par
\medskip
\noindent The main purpose of the present paper is to construct the multiphase solutions of equation (1.1) and investigate their properties in comparison with those of equation (1.2). 
To this end, we employ the direct method (or sometimes called the bilinear transformation method) [16, 17] which
has been used frequently in  obtaining special solutions such as soliton and periodic solutions.  The method of solution is elementary in the sense that it does not require the knowledge of the inverse scattering
transform (IST) and is based only on an elementary theory of determinants. \par
The remaining part of this section addresses a summary of notations.
In section 2, we first transform equation (1.1) to a set of bilinear equations through appropriate dependent variable transformations. The $N$-phase solution   of the bilinear equations is obtained
following the standard procedure of the direct method. It is represented by a quotient of two fundamental determinants (or tau-functions) which play a fundamental role in carrying out the analysis.
To assure the nonsingular nature of the solution, we give a nondegeneracy condition for the matrix associated with a tau-function. This  justifies the assumption in deriving the bilinear equations. 
Subsequently, an alternative  representation of the $N$-phase solution is presented which is analogous to the corresponding one [9] for the $N$-phase solution of equation (1.2).
Last, the eigenvalue problems (1.3)-(1.6) associated with the $N$-phase solution are discussed briefly which would turn out to be a pivot to solving the general initial value problem of equation (1.1).
As a byproduct, we provide another proof of the $N$-phase solution.
In section 3, after parameterizing the $N$-phase solution in terms of wavenumbers and velocities,  the explicit examples of solutions
 are illustrated  for $N=1, 2$ in comparison with the corresponding solutions of equation (1.2).
In section 4, the reduction procedures are performed for the $N$-phase solution.
First, we consider the case  in which all the wavenumbers are set to a positive real constant. The resulting solution is shown to exhibit  a  pole representation whose dynamics  obey the
completely integrable Calogero-Moser-Sutherland dynamical system. The subsequent analysis reveals that the $N$-soliton solution can be  obtained simply 
from the $N$-phase solution by taking the long-wave limit, recovering the
$N$-soliton solution constructed by means of an inverse spectral formula [11]. Then, the one- and two-soliton solutions are displayed, as well as the large-time asymptotic of the general $N$-soliton solution.
Section 5 is devoted to concluding remarks. Most of the technical details are given in appendices. In appendix A, we show that the compatibility  conditions for 
the linear system  (1.3)-(1.6)  indeed yields equation (1.1). As a consequence,
 we derive an infinite number of conservation laws. 
Although they have been obtained by computing the trace of the powers of the Lax operator [11], the present alternative construction uses a linear recursion relation
among the eigenfunctions. 
The equivalency of the Lax equation derived from (1.3)-(1.6) and that given in [11] is demonstrated in (A.3).
In appendix B, we prove various differential rules for the tau-functions. 
In appendix C, the two striking identities are verified for the tau-functions. In appendix D, we provide the proof  of the results for the eigenvalue problems (1.3)-(1.6) presented in section 2.
\par
\bigskip
\noindent 1.4. Notations \par
\medskip
\noindent  We summarize the notations concerning vectors, matrices and bilinear operators which will be used frequently  throughout the paper.
\par
\medskip
\noindent 1)  Row vectors
$${\bf a}=(a_1, a_2, ..., a_N), \quad {\bf b}=(b_1, b_2, ..., b_N),\quad {\bf c}=(c_1, c_2, ..., c_N), \quad {\bf d}=(d_1, d_2, ..., d_N), \eqno(1.7a)$$
$${\bf 1}=\underbrace{(1, 1, ..., 1)}_N,  
\quad  {\bf \hat 1}=\underbrace{(1, 1, ..., 1)}_{N-1}, \quad {\bf e}_1= \underbrace{(1, 0, ..., 0)}_N,  \eqno(1.7b)$$
$${\bf p}=(p_1, p_2, ..., p_N),  \quad {\bf \hat p}=(p_2, p_3, ..., p_N), 
\quad {\bf q}=(q_1, q_2, ..., q_N), \quad {\bf \hat q}=(q_2, q_3. ..., q_N), \eqno(1.7c) $$
$${\bf P}=(p_1^2, p_2^2, ..., p_N^2), \quad  {\bf Q}=(q_1^2, q_2^2, ..., q_N^2), \eqno(1.7d)$$
$${\bf f_1}=( f_{12}, f_{13}, ..., f_{1N}), \quad {\bf f_2}=(f_{21}, f_{31}, ..., f_{N1}), \eqno(1.7e)$$
where $a_j, b_j, c_j, d_j\ (j=1, 2, ..., N)\in \mathbb{C}$, $ f_{1j}, f_{j1} (j=2, 3, ..., N)\in \mathbb{C}$ and $p_j, q_j\ (j=1, 2, ..., N)\in \mathbb{R}$. \par
\noindent 2) Matrices and cofactors\par
$$D=(d_{jk})_{1\leq j ,k\leq N}, \quad D({\bf a}; {\bf b})=\begin{pmatrix} D & {\bf b}^T\\ {\bf a} & 0\end{pmatrix}, $$
$$ D({\bf a}, {\bf b } ; {\bf c}, {\bf d})=\begin{pmatrix} D & {\bf c}^T & {\bf d}^T\\ {\bf a} & 0 &0 \\ {\bf b} & 0 & 0\end{pmatrix}\ ({\rm bordered\ matrices}), \eqno(1.8)$$ 
$$|D|={\rm det}\,D, \quad D_{jk}=\partial |D|/\partial d_{jk}\ ({\rm first\ cofactor\ of}\ d_{jk}), \eqno(1.9a)$$
$$ D_{jk, lm}=\partial^2|D|/\partial d_{jl}\partial d_{km}\ (j<k, l<m)\ ({\rm second\ cofactor}), \eqno(1.9b)$$
where $d_{jk}\ (j, k=1, 2, ..., N)\in \mathbb{C}$ and the symbol $T$ denotes transpose. \par
\medskip
\noindent 3) Bilinear operators \par       
$$D_x^mD_t^ng\cdot f=\left({\partial\over\partial x}-{\partial\over\partial x^\prime}\right)^m
\left({\partial\over\partial t}-{\partial\over\partial t^\prime}\right)^n
g(x, t)f(x^\prime,t^\prime)\Big|_{ x^\prime=x,\,t^\prime=t},\ (m, n=0, 1, 2, ...),  \eqno(1.10a)$$
$$D_tg\cdot f=g_tf-gf_t, \quad D_xg\cdot f=g_xf-gf_x, \quad D^2_xg\cdot f=g_{xx}f-2g_xf_x+gf_{xx}. \eqno(1.10b)$$
\par
\bigskip
\leftline{\bf 2. Multiphase solutions}\par
\medskip
\noindent The goal of this section is to establish the following theorem. \par
\medskip
\noindent{\bf Theorem 2.1.} {\it The $N$-phase solution of equation (1.1) admits a determinantal expression in terms of
 the tau-functions $f=f(x, t)$ and $g=g(x, t)$ 
$$u={g\over f}, \quad f=|F|, \quad g=g_0|G|,\eqno(2.1)$$
where $F$ and $G$ are  $N\times N$ matrices whose elements are given by
$$F=(f_{jk})_{1\leq j,k\leq N}, \quad f_{jk}=\zeta_j\delta_{jk}+{1\over p_j-q_k}, \eqno(2.2a)$$
$$ G=(g_{jk})_{1\leq j,k\leq N}, \quad g_{1k}=1\ (k=1, 2, ..., N), \quad g_{jk}=f_{jk},\quad (j=2, 3, ..., N, k=1, 2, ..., N), \eqno(2.2b)$$
with
$$ g_0=|g_0|\,e^{{\rm i}\chi}, \quad |g_0|^2={1\over q_1}(1-e^{-2q_1\rho})\prod_{j=2}^N{p_j\over q_j}, \eqno(2.3a)$$
$$\zeta_j={e^{-{\rm i}\theta_j+\delta_j}\over p_j-q_j},\quad \theta_j=(p_j-q_j)\left\{x-(p_j+q_j)t-x_{j0}\right\}, \quad (j=1,2, ..., N), \eqno(2.3b)$$
$$\delta_j=\phi_j+{1\over 2}\sum_{\substack{k=1\\ (k\not=j)}}^NA_{jk}, \quad \phi_j=-q_1\rho\delta_{j1},\quad (j=1,2, ..., N), \eqno(2.3c)$$
$$e^{A_{jk}}={(p_j-p_k)(q_j-q_k)\over (p_j-q_k)(q_j-p_k)}, \quad (j, k=1, 2, ..., N; j\not=k). \eqno(2.3d)$$
Here, $\delta_{jk}$ denotes Kronecker's delta, $\rho$ is a positive constant and $x_{j0}\ (j=1, 2, ..., N)\in\mathbb{R}$ and $\chi\in\mathbb{R}$ are phase parameters. 
The real parameters $p_j$ and $q_j$ are imposed on the conditions
$$p_1=0,\quad q_1<0<q_2<p_2< ...<q_N<p_N. \eqno(2.4)$$ }
\par
\noindent {\bf Remark 2.1.} In view of the conditions (2.4), the tau-function $f$ has zeros only in the lower-half complex  plane so that (2.1) gives a nonsingular solution
of equation (1.1) in the upper-half complex  plane. 
The squared modules of $u$ is expressed simply in terms of $f$ and its complex conjugate $f^*$ as
$$|u|^2=-\mu-{\rm i}\,{\partial\over \partial x}\,{\ln}\,{f^*\over f}, \quad \mu=\sum_{j=1}^N(p_j-q_j). \eqno(2.5)$$
Notice that the matrix $G$ is constructed from the matrix $F$ by replacing its first row by the row vector ${\bf 1}$ from $(1.7b)$. 
These statements will be proved later in this section. 
The solutions (2.1) and (2.5) are functions of the $N$ phase variables $\theta_j\ (j=1, 2, ..., N)$ and hence they are called the $N$-phase
solutions. If the parameters $p_j-q_j$ satisfy the conditions $p_j-q_j=2\pi m_j/L\ (j=1, 2, ..., N)$, where $m_j$ are positive integers such that 
$0<m_1\leq m_2\leq ...\leq m_N$ and $L$ is an arbitrary positive constant, then the solutions become periodic functions in the variable $x$. This specific setting
will be considered in section 4.
\par
\bigskip
\leftline{ 2.1. Bilinearization} \par
\medskip
\noindent The proof of theorem 2.1 consists of a sequence of steps.  The first step is to bilinearize equation (1.1) 
in accordance with  the standard procedure of the direct method [16, 17]. \par
\bigskip
\noindent {\bf Proposition 2.1} {\it Through the dependent variable transformations (2.1) and (2.5), equation (1.1) can be
transformed to the set of bilinear equations for $f$ and $g$
$${\rm i}D_tg\cdot f=D_x^2g\cdot f, \eqno(2.6)$$
$${\rm i}D_xf^*\cdot f=-\mu f^*f-g^*g. \eqno(2.7)$$}
\medskip
\noindent {\bf Proof.} Since $f_x/f\,(f^*_x/f^*)$ is analytic in ${\rm Im}\, x\geq 0\ ({\rm Im}\, x \leq 0)$, we deduce  $H(f_x/f)={\rm i}\,f_x/f$
and $H(f^*_x/f^*)=-{\rm i}\,f^*_x/f^*$. This gives
$${1\over 2}(1-{\rm i}H)(|u|^2)_x={\rm i}\,\left({f_x\over f}\right)_x.$$ 
If we substitute the above relation and (2.1) into equation (1.1), we have 
$${\rm i}\left({g\over f}\right)_t=\left({g\over f}\right)_{xx}+2\,{g\over f}\,\left({f_x\over f}\right)_x.$$
We can rewrite this equation in terms of the bilinear operators defined by (1.10) to obtain (2.6). 
The bilinear equation (2.7) follows simply from (2.1) and (2.5). \hspace{\fill}$\Box$ \par
\bigskip
To proceed, we provide the following differential rules for the determinants $|F|$ and $|G|$. \par
\bigskip
\noindent {\bf Lemma 2.1.} \par
$$|F|_t={\rm i}\sum_{j=1}^N(p_j^2-q_j^2)|F|+{\rm i}(|F({\bf 1}; {\bf p})|+|F({\bf q}; {\bf 1})|), \eqno(2.8a)$$
$$|F|_x=-{\rm i}\mu|F|-{\rm i}|F({\bf 1}; {\bf 1})|, \eqno(2.8b)$$
$$|F|_{xx}=-\mu^2|F|-2\mu|F({\bf 1}; {\bf 1})|+|F({\bf 1}; {\bf p})|-|F({\bf q}; {\bf 1})| , \eqno(2.8c)$$
$$|G|_t={\rm i}\sum_{j=1}^N(p_j^2-q_j^2)|G|-{\rm i}|F({\bf Q}; {\bf  e}_1)|-{\rm i}|F({\bf q}, {\bf 1}; {\bf 1}, {\bf  e}_1)|, \eqno(2.9a)$$
$$|G|_x=-{\rm i}\left(\mu|G|-|F({\bf q}; {\bf  e}_1)|\right), \eqno(2.9b)$$
$$|G|_{xx}=-\mu^2|G|+2\mu|F({\bf q}; {\bf  e}_1)|+|F({\bf Q}; {\bf  e}_1)|-|F({\bf q}, {\bf 1}; {\bf 1}, {\bf  e}_1)|. \eqno(2.9c)$$
 \medskip
 The proof of lemma 2.1 is given in appendix B. \par
 \bigskip
 \noindent { 2.2. Proof of theorem 2.1.}\par
 \medskip
 \noindent We now show that the $N$-phase solution given by theorem 1 satisfies the bilinear equations (2.6) and (2.7). The proof is
 carried out by using the basic formulas  of determinants, some of which are summarized in appendix B. Among them, Jacobi's formula (B.3) will play a key
  role. \par
 \medskip
 \noindent { 2.2.1. Proof of (2.6)}\par
 \medskip
 \noindent Define the polynomial $P$ in $\zeta_j\ (j=1, 2, ..., N)$ by
 $$g_0P=D_tg\cdot f-D_x^2g\cdot f=({\rm i}g_t-g_{xx})f-({\rm i}f_t+f_{xx})g+2g_xf_x. $$
 We then substitute (2.8) and (2.9) into $P$ and see that $P$ reduces to the form
 $$P=2|F({\bf q}, {\bf 1}; {\bf 1}, {\bf e}_1)||F|+2|F({\bf q}, {\bf 1})||G|
     +2 |F({\bf q}; {\bf  e}_1)||F({\bf 1}; {\bf 1})|. $$
Using Jacobi's formula (B.3), the first term of $P$ becomes
$$2|F({\bf q}, {\bf 1}; {\bf 1}, {\bf  e}_1)||F|=2\{|F({\bf q}; {\bf 1})||F({\bf 1}; {\bf  e}_1)|-|F({\bf q}; {\bf e}_1)||F({\bf 1}; {\bf 1})|\}. $$
Invoking the definition of the matrix $G$, one  has $|G|=-|F({\bf 1}; {\bf  e}_1)|$.
Hence, the second term can be written as $-2|F({\bf q}; {\bf 1})||F({\bf 1}; {\bf  e}_1)|.$
If we substitute the above two expressions into $P$, $P$ turns out to be zero identically.   This proves (2.6). \hspace{\fill} $\Box$ \par
\medskip
Before proceeding to the proof of (2.7), we prepare the following formulas. \par
\medskip
\noindent {\bf Lemma 2.2.} {\it Let us introduce the determinants $|\bar F|$ and  $|\bar G|$, 
where $\bar F=(\bar f_{jk})_{1\leq j,k\leq N}$ is an $N\times N$ matrix and $\bar G=(\bar g_{jk})_{2\leq j,k\leq N}$ is  an $N-1\times N-1$ matrix whose elements are given respectively by
$$\bar f_{jk}=e^{-2
\phi_j}\zeta_j\delta_{jk}+{1\over p_j-q_k}, \quad \bar g_{jk}={q_j\over p_j}\,\zeta_j\delta_{jk}+{1\over p_j-q_k}. \eqno(2.10)$$
Then, the determinantal identities hold
$$|\bar F|={\rm exp}\left[-\sum_{j=1}^N({\rm i}\theta_j+\phi_j)\right]\,|F|^*, \eqno(2.11)$$
$$|\bar G|={\rm exp}\left[-{\rm i}\sum_{j=2}^N\theta_j-{1\over 2}\sum_{k=2}^NA_{1k}\right]\,|G|^*. \eqno(2.12)$$}
 The proof of lemma 2.2 is given in appendix C. \par
 \medskip
 \noindent { 2.2.2. Proof of (2.7)}\par
 \medskip
\noindent  Let $Q={\rm i}D_xf^*\cdot f +\mu f^*f+g^*g$ and $\bar f=|\bar F|$. Thanks to (2.11), one can rewrite $f^*$ in terms of $\bar f$ to obtain
 $${\rm i}D_xf^*\cdot f +\mu f^*f={\rm i}\,{\rm exp}\left[\sum_{j=1}^N({\rm i}\theta_j+\phi_j)\right]D_x\bar f\cdot f.\eqno(2.13) $$
 \par
 Now, using the differential rule (B.1) and an analogous one
 $${\bar f}_x=-{\rm i}\mu|\bar F|-{\rm i}|\bar F({\bf 1}; {\bf 1})|,$$
we deduce
$${\rm i}D_x\bar f\cdot f=|\bar F({\bf 1}; {\bf 1})||F|-| F({\bf 1}; {\bf 1})||\bar F|. \eqno(2.14)$$
Since the $(1, 1)$ element of the matrix $\bar F({\bf 1}; {\bf 1})$ can be written as $e^{-2\phi_1}\zeta_1+1=\zeta_1+1+(e^{-2\phi_1}-1)\zeta_1$, an elementary 
manipulation yields the relation
$$|\bar F({\bf 1}; {\bf 1})|=|F({\bf 1}; {\bf 1})|+(e^{-2\phi_1}-1)\zeta_1|\hat F({\bf \hat 1}; {\bf \hat 1 })|. $$ 
Similarly, one has
$$|\bar F|=|F|+(e^{-2\phi_1}-1)\zeta_1|\hat F|, $$
where $\hat F=(f_{jk})_{2\leq j,k\leq N}$ is an $N-1\times N-1$ matrix.
Substituting these two expressions into (2.14), we recast it to the form
$${\rm i}D_x\bar f\cdot f=\left(e^{-2\phi_1}-1\right)\zeta_1(Q_1-Q_2), \eqno(2.15a)$$
with
$$Q_1=|\hat F({\bf \hat 1}; {\bf \hat 1 })||F|, \quad Q_2=|F({\bf 1}; {\bf 1})||\hat F|. \eqno(2.15b)$$
\par
We modify $|F({\bf 1}; {\bf 1})|$ and $|F|$ with the aid of the basic rules for determinants as
$$|F({\bf 1}; {\bf 1})|=|\hat F({\bf \hat 1}; {\bf \hat 1 })|f_{11}-(|\hat F({\bf \hat 1}; {\bf f}_2|+|\hat F|)-|\hat F({\bf f}_1; {\bf \hat 1 })|+|\hat F({\bf f}_1,{\bf\hat 1}; {\bf f}_2,{\bf \hat 1})|, $$
$$|F|=|\hat F|f_{11}+|\hat F({\bf f}_1; {\bf f}_2)|,$$
where ${\bf f}_1$ and ${\bf f}_2$ are row vectors defined by (1.7e). 
By using these expressions  as well as Jacobi's formula
$$|\hat F({\bf f}_1,{\bf \hat 1}; {\bf f}_2,{\bf \hat 1})||\hat F|=|\hat F({\bf f}_1; {\bf f}_2)||\hat F({\bf \hat 1}; {\bf \hat 1 })|-|\hat F({\bf f}_1; {\bf \hat 1 })||\hat F({\bf  \hat 1}; {\bf f}_2)|,$$
and  an identity  $|\hat F({\bf \hat 1}; {\bf f}_2)|=|G|-|\hat F|$, we  obtain
\begin{align}
Q_1-Q_2 &=(|\hat F|+|\hat F({\bf f}_1; {\bf \hat 1 })|)|G| \notag \\
        &= \prod_{j=2}^N{p_j\over q_j}\,|\bar G||G|. \tag{2.16}
        \end{align}
Note that the second line of (2.16) comes from (C.6).
It now follows from (2.13), (2.15) and (2.16) that 
$$Q={\rm exp}\left[\sum_{j=1}^N({\rm i}\theta_j+\phi_j)\right]\,\left(e^{-2\phi_1}-1\right)\zeta_1\prod_{j=2}^N{p_j\over q_j}\,|\bar G||G|+|g_0|^2|G|^*|G|.$$
We introduce $|\bar G|$ from (2.12) 
 and use the relations 
$$\zeta_1=-{1\over q_1}\,{\rm exp}\left[-{\rm i}\theta_1-{1\over 2}\sum_{k=2}^NA_{1k}+\phi_1\right],\quad  \phi_1=-q_1\rho, \quad\phi_j=0\quad j\geq 2,$$
  to reduce the quantity $Q$ to the form
$$Q=\left\{-{1\over q_1}\left(1-e^{-2q_1\rho}\right)\prod_{j=2}^N{p_j\over q_j}+|g_0|^2\right\}|G|^*|G|. $$
In view of (2.3a), we finally arrive at the identity $Q=0$, completing the proof of (2.7). \hspace{\fill}$\Box$ \par
\bigskip
\leftline{2.3. Analyticity of the tau-function $f$} \par
\medskip
\noindent The basic assumption in deriving the bilinear equations (2.6) and (2.7) is that the tau-function $f$ has no zeros in the upper-half complex plane.
 The proposition below describes this statement clearly. \par
\medskip
\noindent {\bf Proposition 2.2.}\ {\it The matrix $F$ is nondegenerate in ${\rm Im}\,x\geq 0$ provided that the conditions (2.4) are satisfied. }\par
\medskip
\noindent {\bf Proof.} We follow the argument developed in [1, 18] for the multiphase solutions of the Benjamin-Ono (BO) equation. 
First, we put $f_{jk}=\zeta_j\left(\delta_{jk}+{\alpha_j\over p_j-q_k}\right)\equiv \zeta_ja_{jk}$ with $\alpha_j=\zeta_j^{-1}$ and consider the matrix  $A=(a_{jk})_{1\leq j,k\leq N}$.
Assume that $A$ is generate for some $x=x_0\in \mathbb{C}$ and $t\in \mathbb{R}$. It implies that the system of linear algebraic equations for $r_j\in\mathbb{C}$ 
$$r_j+\sum_{k=1}^N{\alpha_j\over p_j-q_k}\,r_k=0, \quad (j= 1, 2, ..., N), \eqno(2.17)$$
has a nonzero solution.
Define a meromorphic function $\psi(z)=\sum_{k=1}^N{r_k\over z-q_k}. $ Then, the system of equations (2.17) can be written as $N$ conditions
$r_j=-\alpha_j\psi(p_j) \ (j=1, 2, ..., N)$. \par
We introduce the function $R$ by
$$R(z)=\psi(z)\psi^*(z^*)\prod_{k=1}^N{z-q_k\over z-p_k}, \eqno (2.18)$$
where $\psi^*(z^*)=\sum_{k=1}^N{r_k^*\over z-q_k}$.
Let $\mathscr{R}$ be the sum of residues of $R(z)$ at $z=p_j, z=q_j\ (j=1, 2, ..., N)$. By a simple computation, we find that 
$$\mathscr{R}=\sum_{j=1}^N(p_j-q_j){|r_j|^2\over |\alpha_j|^2}\prod_{\substack{k=1\\ (k\not=j)}}^N{p_j-q_k\over p_j-p_k}
\left\{1-{|\alpha_j|^2\over (p_j-q_j)^2}\prod_{\substack{k=1\\ (k\not=j)}}^N{(p_j-p_k)(q_j-q_k)\over (p_j-q_k)(q_j-p_k)}\right\}.\eqno(2.19a) $$
Using the relation
$$|\alpha_j|^2=(p_j-q_j)^2{\rm exp}\left[-2(p_j-q_j)\,{\rm Im}\, x_0-\sum_{\substack{k=1\\ (k\not=j)}}^NA_{jk}+2\rho q_1\delta_{j1}\right], $$
which follows from (2.2a) as well as the condition $p_1=0$, $\mathscr{R}$ recasts to
$$\mathscr{R}=-q_1{|r_1|^2\over |\alpha_1|^2}\left(1-e^{2q_1({\rm Im}\,x_0+\rho}\right)\prod_{k=2}^N{q_k\over p_k}
+\sum_{j=2}^N(p_j-q_j){|r_j|^2\over |\alpha_j|^2}\left(1-e^{-2(p_j-q_j)\,{\rm Im}\,x_0}\right)\prod_{\substack{k=1\\ (k\not=j)}}^N{p_j-q_k\over p_j-p_k}.\eqno(2.19b) $$
On the other hand, since $R(z) \sim z^{-2}$ as $z\rightarrow \infty$, evaluating the residue at $z=\infty$ gives $\mathscr{R}=0$.
To compute $\mathscr{R}$ from (2.19), one has two scenarios according to  values of ${\rm Im}\,x_0$. If ${\rm Im}\,x_0>0$, then $\mathscr{R}>0$ by (2.4) and (2.19b) with $\rho>0, q_1<0$, which is a contradiction.
Hence, $r_j=0\ (j=1, 2, ..., N)$, implying that the matrix $A$ is nondegenerate, or equivalently 
 $|F|=\prod_{j=1}^N\zeta_j|A|\not=0$ for ${\rm Im}\,x_0>0$. \par
 We consider the second case ${\rm Im}\,x_0=0$. It then turns out from (2.19b) that
$\mathscr{R}=-q_1|r_1|^2/|\alpha_1|^2(1-e^{2q_1\rho})\prod_{k=2}^N(q_k/p_k)\geq 0$. Hence, $\mathscr{R}$ becomes zero only if $r_1=0$,  which we address in detail. 
When $r_1=0$, the system of equations (2.17) becomes
$$\sum_{k=2}^N{r_k\over q_k}=0, \quad r_j+\sum_{k=2}^N{\alpha_j\over p_j-q_k}\,r_k=0, \quad (j= 2, 3,  ..., N). \eqno(2.20)$$
Eliminating $r_N$ by means of the first equation of (2.20), one can rewrite the second equation  in the form
$$\tilde r_j+\sum_{k=2}^{N-1}{\tilde \alpha_j\over p_j-q_k}\tilde r_k=0, \quad (j= 2, 3, ..., N-1), \eqno(2.21a)$$
with
$$\tilde r_j={q_j-q_N\over q_j}\,r_j, \quad \tilde \alpha_j=\zeta_j^{-1}\,{p_j(q_j-q_N)\over q_j(p_j-q_N)},\quad (j= 2,3,  ..., N-1). \eqno(2.21b)$$
We then find  the expression $\tilde{\mathscr{R}}$ of the residue corresponding to (2.19a) 
$$\tilde{\mathscr{R}}=\sum_{j=2}^{N-1}(p_j-q_j){|\tilde r_j|^2\over |\tilde\alpha_j|^2}\prod_{\substack{k=2\\ (k\not=j)}}^{N-1}{p_j-q_k\over p_j-p_k}
\left\{1-{|\tilde\alpha_j|^2\over (p_j-q_j)^2}\prod_{\substack{k=2\\ (k\not=j)}}^{N-1}{(p_j-p_k)(q_j-q_k)\over (p_j-q_k)(q_j-p_k)}\right\}. \eqno(2.22) $$
Since ${\rm Im}\,x_0=0$, it follows from (2.3b) that 
$$|\zeta_j|^{-2}=(p_j-q_j)^2\,{\rm exp}\left[-\sum_{\substack{k=1\\ (k\not=j)}}^NA_{jk}\right], \quad (j=2, 3, ..., N-1). $$ 
By using this expression to evaluate $|\tilde\alpha_j|^2$, $\tilde{\mathscr{R}}$ becomes
$$\tilde{\mathscr{R}}=-\sum_{j=2}^{N-1}(p_j-q_j)^2{|\tilde r_j|^2\over |\tilde\alpha_j|^2}{(p_j+q_j-q_1)p_Nq_N-(p_N+q_N-q_1)p_jq_j\over q_j(p_N-p_j)(q_N-p_j)(q_j-q_1)}
\prod_{\substack{k=2\\ (k\not=j)}}^{N-1}{(p_j-q_k)\over (p_j-p_k)}.$$
We add the inequalities $1/p_j>1/p_N, 1/q_j>1/q_N, -q_1/(p_jq_j)>-q_1/(p_Nq_N)\ (q_1<0,\ p_j,\, q_j>0\ (j=2, 3, ..., N-1))$ together and take into account  (2.4) to obtain the inequality
$$(p_j+q_j-q_1)p_Nq_N-(p_N+q_N-q_1)p_jq_j$$
$$=p_jq_jp_Nq_N\left\{{1\over p_j}+{1\over q_j}-{q_1\over p_jq_j}-\left({1\over p_N}+{1\over q_N}-{q_1\over p_Nq_N}\right)\right\}>0. $$
It turns out from the above observation  and (2.4) that $\tilde{\mathscr{R}}<0$. Since $\tilde{\mathscr{R}}=0$, this leads to a contradiction. Consequently,
$\tilde r_j=0\ (j=2, 3, ..., N-1)$, which yields $\tilde r_N=0$ by virtue of the first equation  of (2.20).  
When we plug this result into   the condition $\tilde r_1=0$, we find $\tilde r_j=0\ (j=1, 2, ..., N)$ and hence $|F|\not=0$ for ${\rm Im}\, x_0=0$.
Consequently, $f=|F|\not=0$ for ${\rm Im}\,x_0\geq 0.$  Thus, we finish the proof of  proposition 2.2.   \hspace{\fill}$\Box$ \par
\medskip
\noindent {\bf Remark 2.2.}\ The tau-function $f$ has zeros in the lower-half complex plane. Actually, it can be inferred from (2.19b) that $\mathscr{R}$ may become zero in the range of the parameter
$-\rho<{\rm Im}\,x_0<0$.  Although a number of  numerical computations confirm this conjecture, its  verification needs the rigorous analysis. \par
\bigskip
\noindent {2.4. Alternative representation of the $N$-phase solution}\par
\medskip
\noindent As in the case of the $N$-phase solution of equation (1.2) [3, 9], the $N$-phase solution constructed in theorem 2.1 has an alternative representation
which provides an analog of Dubrobin's formulation for the multiphase solutions of the Korteweg-de Vries equation [19]. We describe it by the
following proposition. \par
\medskip
\noindent {\bf Proposition 2.3.}\ {\it The squared modules of the $N$-phase solution of equation (1.1) admits a representation
$$|u|^2=\sum_{j=1}^N(p_j+q_j)-{\rm i}\,\sum_{j=1}^N(\mu_j-\mu_j^*), \eqno(2.23)$$
where the functions $\mu_j=\mu_j(x, t)$ solve the system of nonlinear algebraic equations
$${\prod_{\substack{k=1\\ (k\not=j)}}^N(q_j-q_k)\over \prod_{k=1}^N(p_j-q_k)}
{\prod_{k=1}^N(p_j-{\rm i}\mu_k)\over \prod_{k=1}^N(q_j-{\rm i}\mu_k)}=-\zeta_j,\quad (j=1, 2, ..., N). \eqno(2.24)$$
They also satisfy the system of nonlinear PDEs
$$\sum_{k=1}^N{\mu_{k,x}\over (p_j-{\rm i}\mu_k)(q_j-{\rm i}\mu_k)}=-1, \quad \sum_{k=1}^N{\mu_{k,t}\over (p_j-{\rm i}\mu_k)(q_j-{\rm i}\mu_k)}=p_j+q_j,\quad (j=1, 2, ..., N). \eqno(2.25)$$}
\par
\bigskip
\noindent{\bf Proof.}\  First, consider the system of
linear algebraic equations ${\rm i}\sum_{j=1}^Nf_{jk}\Phi_k=1$ for $\Phi_j$ \ $(j=1,2, ..., N)$. By Cramer's rule, the solution is found to be
$$\Phi_j=-{\rm i}\,{\sum_{k=1}^NF_{kj}\over |F|}, \quad (j=1, 2, ..., N). \eqno(2.26)$$
Referring to formula (2.8b), we obtain
$${\rm i}\,\sum_{j=1}^N\,\Phi_j=\sum_{j=1}^N(p_j-q_j)-{\rm i}\,{|F|_x\over |F|}, $$
which, plugged into its complex conjugate expression, gives
$${\rm i}\,\sum_{j=1}^N\,(\Phi_j-\Phi_j^*)=2\sum_{j=1}^N(p_j-q_j)-{\rm i}\, {\partial\over \partial x}\,\ln\,{|F|\over |F|^*}. $$
Using this relation, one can rewrite $|u|^2$ from (2.5) in the form
$$|u|^2=\sum_{j=1}^N(p_j-q_j)-{\rm i}\,\sum_{j=1}^N\,(\Phi_j-\Phi_j^*). \eqno(2.27) $$
\par
Introduce the functions $G=G(x, t, \lambda)$ and $\mu_j=\mu_j(x, t)\ (j=1, 2, ..., N)$ by
\begin{align}
G(x, t, \lambda)&=1-{\rm i}\,\sum_{j=1}^N{\Phi_j\over \lambda-q_j} \tag{2.28a} \\
                &={\prod_{j=1}^N(\lambda-{\rm i}\mu_j)\over \prod_{j=1}^N(\lambda-q_j)}.\tag{2.28b} 
\end{align}
It follows from (2.28a) by using the equation for $\Phi_j$ that
$$G(x, t, p_j)={\rm i}\zeta_j\,\Phi_j, \eqno(2.29)$$
 and from (2.28b) with $\lambda=p_j$ that
 $$G(x, t, p_j)={\prod_{k=1}^N(p_j-{\rm i}\mu_k)\over \prod_{k=1}^N(p_j-q_k)}. \eqno(2.30)$$
 On the other hand, by computing the residue at $\lambda=q_j$ in (2.28), we find
 $$\Phi_j={\rm i}\,{\prod_{k=1}^{N}(q_j-{\rm i}\mu_k)\over \prod_{\substack{k=1 \\ (k\not= j)}}^{N}(q_j-q_k)}. \eqno(2.31)$$
 If we equate (2.29) with (2.30) and then insert $\Phi_j$ from (2.31) into the resultant expression, we establish (2.24).
 The system of PDEs (2.25) for $\mu_j$  follows by taking the logarithmic derivative of (2.24) 
 with respect to $x$ and $t$, respectively. Expanding (2.28a) and (2.28b) in inverse powers of $\lambda$ and comparing the coefficient of $\lambda^{-1}$, we
 obtain the relation $\sum_{j=1}^N\Phi_j=\sum_{j=1}^N(\mu_j+{\rm i}q_j)$. Introducing this expression and its complex conjugate into (2.27) yields (2.23).  \hspace{\fill}$\Box$
  \par
  \bigskip
 \noindent{2.5. Eigenvalue problem for the $N$-phase solution} \par
 \medskip
 \noindent  In developing the formulation of IST for equation (1.1), the spectral analysis of the 
 spatial part of  the associated eigenvalue equations plays a central role. Here,
 we demonstrate shortly  that the eigenvalue equations (1.3), (1.5) and (1.6) can be solved exactly for the $N$-phase solution (or time dependent $N$-phase potential).
 A full analysis for the generic potentials will be devoted to a future work. See [7] for an analogous 
work on the eigenvalue problem associated with the $N$-soliton solution of equation (1.2).
The following proposition describes the main result. \par
\medskip
\noindent {\bf Proposition 2.4.} {\it Let $\phi_ j, \psi_j^+$  and $\lambda_j$ be the eigenfunctions and corresponding eigenvalue
for equation (1.3) with $u$ being the $N$-phase solution (2.1). If  $\phi_ j$ and $ \psi_j^+$  satisfy the system of linear algebraic equations
$$\sum_{k=1}^Nf_{jk}\phi_k=g_0\delta_{j1}, \quad (j=1, 2, ..., N), \eqno(2.32a) $$
$$ \sum_{k=1}^Nf_{jk}\psi_k^+=-1, \quad (j=1, 2, ..., N), \eqno(2.32b)$$
then they solve the eigenvalue equations
$${\rm i}\phi_{j,x}+\lambda_j\phi_j+u\psi_j^+=0,\quad (j=1, 2, ..., N), \eqno(2.33)$$
$${\rm i}\phi_{j,t}-2{\rm i}\lambda_j \phi_{j,x}+\phi_{j,xx}-2{\rm i}u_x\psi_j^+=0, \quad (j=1, 2, ..., N),\eqno(2.34)$$
$${\rm i}\psi^+_{j,t}-2{\rm i}\lambda_j \psi^+_{j,x}+\psi^+_{j,xx}-{\rm i}[(1-{\rm i}H)(|u|^2)_x]\psi_j^+=0, \quad (j=1, 2, ..., N),\eqno(2.35)$$
with $\lambda_j=-q_j\ (j=1, 2, ..., N)$.}
\par
\medskip
The proof of proposition 2.4 is given in appendix D. As will be shown in Appendix A, the compatibility conditions of equations
(2.33)-(2.35) yield equation (1.1). Hence, the proposition above provides an alternative proof of the $N$-phase solution.
\par
\bigskip
\leftline{\bf 3. Explicit examples} \par
\medskip
\noindent 
\noindent {3.1. New parameterization of the $N$-phase solution}\par
\medskip
\noindent The $N$-phase solution (2.1) is characterized by the parameters $\rho, \chi, p_j, q_j$ and $x_{j0}\ (j=1, 2, ..., N)$.  
The parameters $p_j$ and  $q_j$ are constrained by the conditions (2.4) to assure the analyticity of the solution. In order to clarify
the physical implication of the solution, it is useful to introduce the wavenumber $k_j$ and the velocity $v_j$ according to the relations
$$k_j=p_j-q_j, \quad v_j=p_j+q_j, \quad (j=1, 2, ..., N). \eqno(3.1)$$
Then, the expressions (2.1)-(2.3) and (2.5)  can be  rewritten in the form
$$u={g\over f}, \quad |u|^2=-\sum_{j=1}^Nk_j-{\rm i}\,{\partial\over \partial x}\,\ln {f^*\over f}, \quad f=|F|, \quad  g=g_0|G|, \eqno(3.2)$$
where
$$\quad F=(f_{jl})_{1\leq j,l\leq N}, \quad f_{jl}=\zeta_j\delta_{jl}+{2\over v_j-v_l+k_j+k_l}, \eqno(3.3a)$$
$$ G=(g_{jk})_{1\leq j,k\leq N}, \quad g_{1k}=1\ (k=1, 2, ..., N), \quad g_{jk}=f_{jk},\quad (j=2, 3, ..., N, k=1, 2, ..., N), \eqno(3.3b)$$
with
$$\quad g_0=|g_0|\,e^{{\rm i}\chi}, \quad |g_0|^2={1-e^{-2k_1\rho}\over k_1}\prod_{j=2}^N{v_j+k_j\over v_j-k_j}, \eqno(3.4a)$$
$$\zeta_j={e^{-{\rm i}\theta_j+\delta_j}\over k_j}={e^{-{\rm i}k_j\xi_j+\delta_j}\over k_j}, \quad \xi_j=x-v_jt-x_{j0}, \quad (j=1, 2, ..., N), \eqno(3.4b)$$
$$\delta_j=\phi_j+{1\over 2}\sum_{\substack{k=1\\ (k\not=j)}}^NA_{jk}, \quad \phi_j=k_1\rho\delta_{j1},\quad (j=1,2, ..., N), \eqno(3.4c)$$
$$e^{A_{jl}}={(v_j-v_l)^2-(k_j-k_l)^2\over (v_j-v_l)^2-(k_j+k_l)^2}, \quad (j, l=1, 2, ..., N; j\not=l). \eqno(3.4d)$$
The conditions (2.4) become
$$k_j>0,\ (j=1, 2, ..., N), \quad v_1<0,\quad v_j>0, \quad v_j>k_j, \ (2\leq j\leq N). \eqno(3.5)$$
\par
\medskip
\noindent{\bf Remark 3.1.}\ With the parameterization (3.1), the $N$-phase solution of equation (1.2) can be written in the form [3, 9]
$$u={g\over f}, \quad |u|^2=1+\sum_{j=1}^Nk_j+{\rm i}\,{\partial\over \partial x}\,\ln {f^*\over f},\quad  f=|F|, \quad g=g_0|G|,\eqno(3.6)$$
where
$$\quad F=(f_{jl})_{1\leq j,l\leq N}, \quad f_{jl}= \tilde\zeta_j \delta_{jl}+{2\over v_j-v_l+k_j+k_l}, \eqno(3.7a)$$
$$G=(g_{jl})_{1\leq j,l\leq N},\quad \quad g_{jl}=e^{\tilde\psi_j-\tilde\phi_j}\tilde\zeta_j \delta_{jl}+{2\over v_j-v_l+k_j+k_l},  \eqno(3.7b)$$
with
$$\quad g_0=|g_0|\,e^{{\rm i}\chi}, \quad |g_0|^2=\prod_{j=1}^N{v_j-k_j\over v_j+k_j}, \eqno(3.8a)$$
$$\tilde\zeta_j={e^{-{\rm i}\theta_j+\tilde\delta_j}\over k_j}={e^{-{\rm i}k_j\xi_j+\tilde\delta_j}\over k_j}, \quad \xi_j=x-v_jt-x_{j0}, \quad (j=1, 2, ..., N), \eqno(3.8b)$$
$$\tilde\delta_j=\tilde\phi_j+{1\over 2}\sum_{\substack{k=1\\ (k\not=j)}}^N\tilde A_{jk}, \quad
e^{2\tilde\phi_j}={(v_j+k_j+2)(v_j-k_j)\over (v_j-k_j+2)(v_j+k_j)},
\quad e^{\tilde\psi_j}={v_j+k_j\over v_j-k_j}\,e^{\tilde\phi_j},\quad (j=1, 2, ..., N), \eqno(3.8c)$$
$$e^{\tilde A_{jl}}={(v_j-v_l)^2-(k_j-k_l)^2\over (v_j-v_l)^2-(k_j+k_l)^2}, \quad (j, l=1, 2, ..., N; j\not=l), \eqno(3.8d)$$
and the parameters $k_j$ and $v_j$ are imposed on  the conditions
$$k_j>0, \quad v_j+k_j<0, \quad v_j-k_j+2>0, \quad (j=1, 2, ..., N). \eqno(3.9)$$
\bigskip
\noindent {3.2. Examples of solutions}\par
\medskip
\noindent {3.2.1. One-phase solution}\par
\medskip
\noindent The one-phase solution is the fundamental constituent among the class of periodic solutions. It reads in the form
$$u={g_0\over e^{-{\rm i}k_1\xi_i+k_1\rho}+1}, \eqno(3.10a)$$
$$|u|^2={k_1\sinh\,k_1\rho\over \cos\,k_1\xi_1+\cosh\,k_1\rho}, \eqno(3.10b)$$
where
$$ g_0=|g_0|\,e^{{\rm i}\chi}, \quad |g_0|^2=(e^{2k_1\rho}-1)/k_1, \quad (k_1>0, \ \rho>0), \quad  \xi_1=x-v_1t-x_{10}. \eqno(3.10c)$$
The one-phase solution represents a nonlinear periodic traveling wave with the period $2\pi/k_1$. The tau-function $f$ associated with it has zeros only in 
the lower-half complex  plane whose imaginary part is given by  $-\rho$. See remark 2.2. \par
It is instructive to compare the one-phase solution with
the corresponding one for the defocusing nonlocal NLS equation (1.2). Explicitly, it follows from (3.6) and (3.7) with $N=1$ that
$$u=g_0\,{e^{-{\rm i}k_1\xi_1+\tilde\psi_1}+1\over e^{-{\rm i}k_1\xi_1+\tilde\phi_1}+1}, \eqno(3.11a)$$
$$|u|^2=1-{k_1\sinh\,\tilde\phi_1\over \cos\,k_1\xi_1+\cosh\,\tilde\phi_1}, \eqno(3.11b)$$
where
$$g_0=|g_0|\,e^{{\rm i}\chi},\quad |g_0|^2={v_1-k_1\over v_2-k_2}, \quad \xi_1=x-v_1t-x_{10}, \eqno(3.11c)  $$
$$ e^{2\tilde\phi_1}={(v_1+k_1+2)(v_1-k_1)\over (v_1-k_1+2)(v_1+k_1)},\quad e^{\tilde\psi_1}={v_1+k_1\over v_1-k_1}\,{\tilde\phi_1}. \eqno(3.11d)$$
Note that in accordance with  (3.9),  the conditions $k_1>0, v_1+k_1<0$ and $v_1-k_1+2>0$ must be imposed on the parameters $k_1$ and $v_1$.
We can see that while the allowable values of the velocity in the expression of  the solution (3.10) lie in the interval $-\infty<v_1<0$ as indicated by (3.5),
the corresponding velocity in the solution (3.11) has an upper limit $|v_1|=2$.  Actually, its value  enters into the region surrounded by the three straight lines
$k_1=0, v_1=-k_1, v_1=k_1-2$ in the $(k_1, v_1)$ plane. It is interesting to observe that  when $v_1=-1$, the minimum value of $|u|^2$ evaluated by (3.11b) becomes zero
regardless of the value of $k_1$. Figure 1 plots the profiles of the modulus  $|u|$ for (3.10b) and (3.11b).
The parameters are set to $\rho=1, k_1=1/2$ for (3.10b) and $k_1=1,  v_1=-3/4$ for (3.11b).  
\bigskip
\begin{center}
\includegraphics[width=8cm]{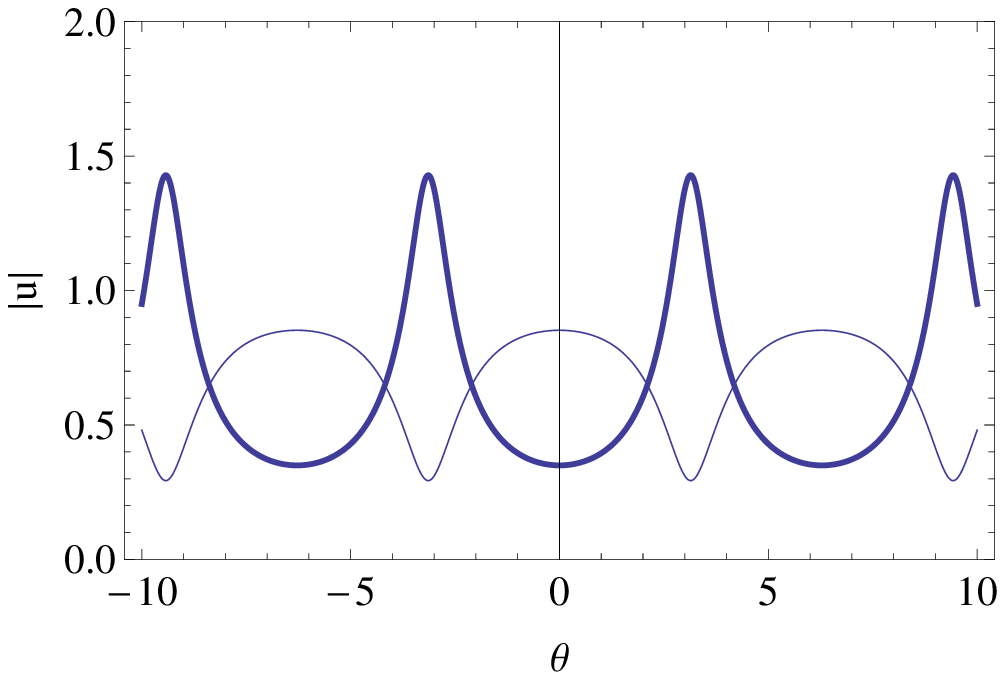}
\end{center}
\noindent  {\bf Figure 1.}\ The profiles of the modulus  $|u|$ are plotted for the one-phase solutions (3.10b) (thick curve) and (3.11b) (thin curve) as function of the phase variable $\theta(=k_1\xi_1)$.
\par
\bigskip
\noindent {3.2.2. Two-phase solution}\par
\medskip
\noindent The two-phase solution from (3.2) with $N=2$ can be written in the form
$$u=g_0\,{|G|\over |F|}, \eqno(3.12a)$$
$$|u|^2=-(k_1+k_2)-{\rm i}\,{\partial\over \partial x}\,{\rm ln}\left(|F|^*\over |F|\right), \eqno(3.12b)$$
with
$$|F|=\zeta_1\zeta_2+{\zeta_1\over k_2}+{\zeta_2\over k_1}+{1\over k_1k_2}{(v_1-v_2)^2-(k_1-k_2)^2\over (v_1-v_2)^2-(k_1+k_2)^2}, \eqno(3.13a)$$
$$|G|=\zeta_2+{1\over k_2}{v_1-v_2+k_1-k_2\over v_2-v_1+k_1+k_2}, \eqno(3.13b)$$
$$g_0=|g_0|\,e^{{\rm i}\chi},\quad |g_0|^2={v_2+k_2\over k_1(v_2-k_2)}\,\left(e^{2k_1\rho}-1\right). \eqno(3.13c)$$
\par
\begin{figure}[h]
\begin{minipage}{0.22\textwidth}
\includegraphics[width=\textwidth]{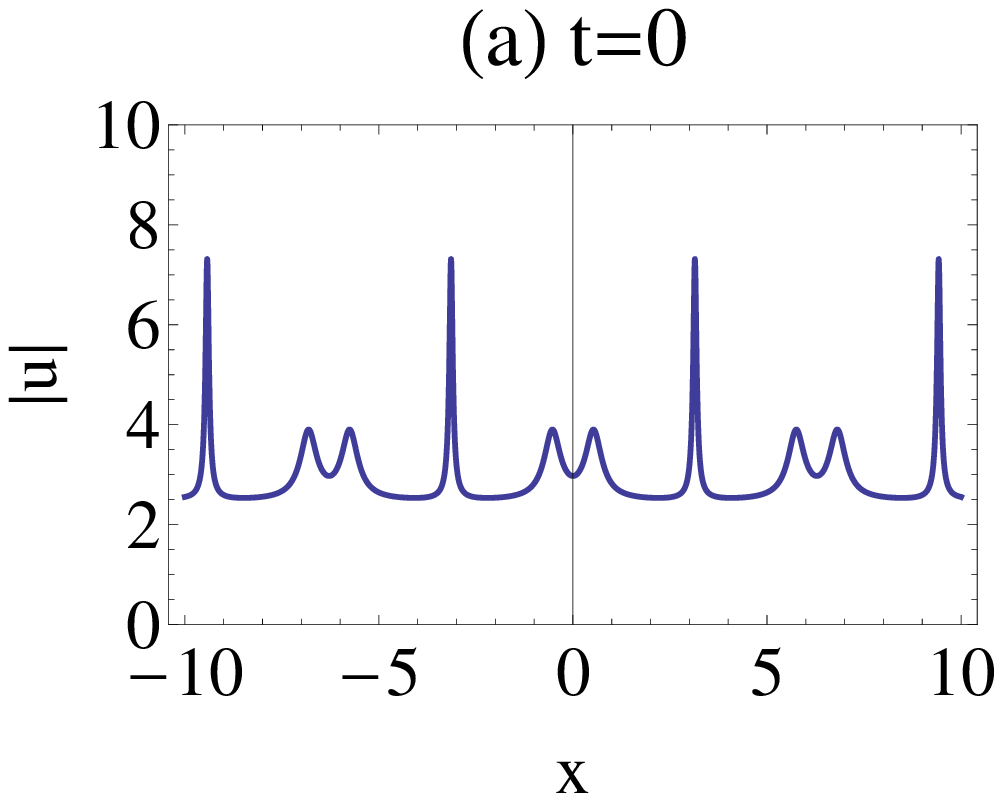}
\end{minipage}
\hspace{9.6pt}
\begin{minipage}{0.22\textwidth}
\includegraphics[width=\textwidth]{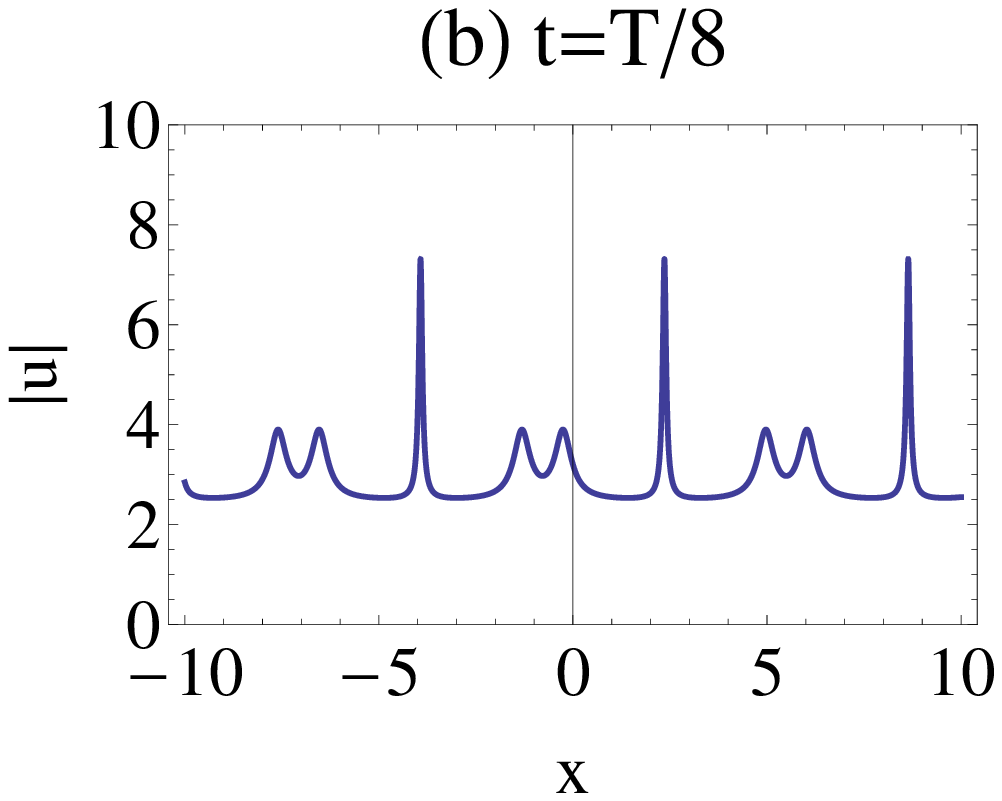}
\end{minipage}
\hspace{9.6pt}
\begin{minipage}{0.22\textwidth}
\includegraphics[width=\textwidth]{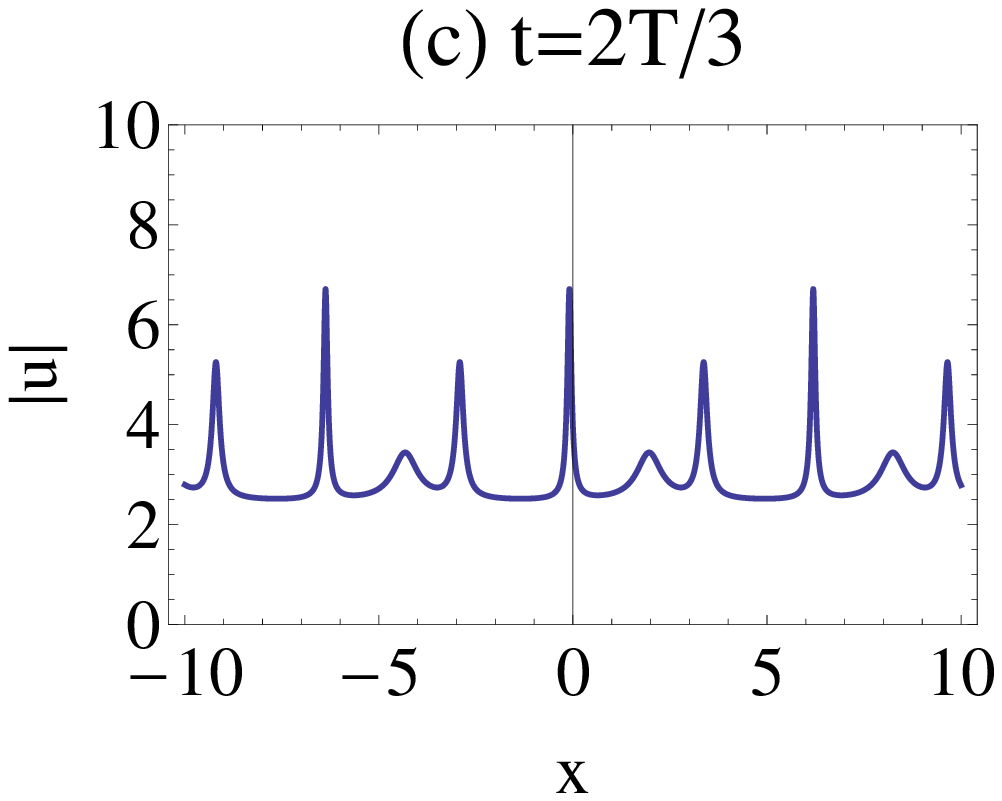}
\end{minipage}
\hspace{9.6pt}
\begin{minipage}{0.22\textwidth}
\includegraphics[width=\textwidth]{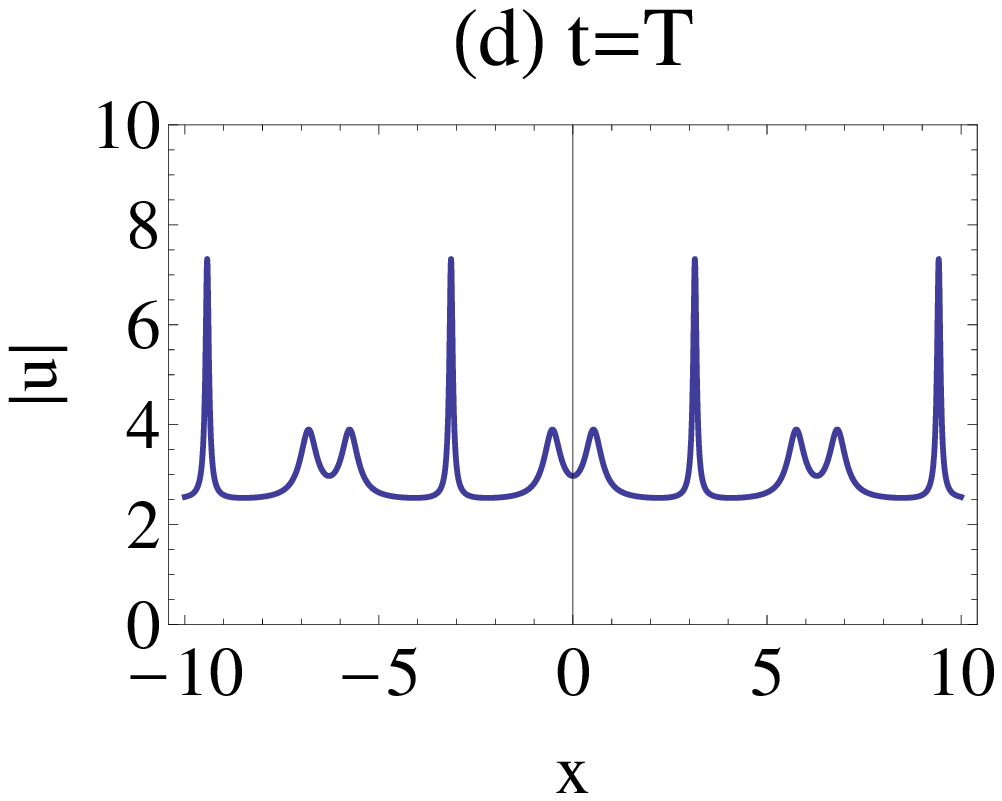}
\end{minipage}
\noindent  {\bf Figure 2}\ The profiles of the modulus  $|u|$ are plotted for the two-phase solution (3.12b)  as function of $x$.
The parameters are set to $\rho=1, k_1=1, k_2=2, v_1=-1, v_2=3, x_{10}=x_{20}=0$.  The period  is $T=2\pi$.  The recurrence time is $T_R=\pi/4=T/8$.\par
\end{figure}

\bigskip
Figure 2 depicts the profiles of $|u|$ from (3.12b) for four different times. The two-phase solution consists of two waves interacting with each other repeatedly.
The period of the slow wave is given by $T_1=2\pi/(k_1|v_1|)=2\pi$ whereas that of the fast one is $T_2=2\pi/(k_2|v_2|)=\pi/3$.
As a result,  the initial profile is recovered at $t=T_1=2\pi$. 
One can see that at a  time $t=T_R=2\pi(k_2-k_1)/\{k_1k_2|v_2-v_1|\}=\pi/4$, the initial profile is recovered, but  with a  phase shift.
This formula for the recurrence time $T_R$ can be derived by eliminating the spatial coordinate $x$ from the relations $\theta_1=2\pi, \theta_2=2\pi$ with $x_{10}=x_{20}=0$.
\par
The two-phase solution of equation (1.2)  is given by (3.6) with $N=2$. It reads in the form
$$u=g_0\,{|G|\over |F|}, \eqno(3.14a)$$
$$|u|^2=1+k_1+k_2+{\rm i}\,{\partial\over \partial x}\,{\rm ln}\left(|F|^*\over |F|\right), \eqno(3.14b)$$
with
$$|F|= \tilde\zeta_1\tilde\zeta_2+{\tilde\zeta_1\over k_2}+{\tilde\zeta_2\over k_1}+{1\over k_1k_2}{(v_1-v_2)^2-(k_1-k_2)^2\over (v_1-v_2)^2-(k_1+k_2)^2}, \eqno(3.15a)$$
$$|G|={(v_1+k_1)(v_2+k_2)\over(v_1-k_1)(v_2-k_2)}\, {\tilde\zeta_1\tilde\zeta_2}+{v_1+k_1\over v_1-k_1}\,{\tilde\zeta_1\over k_2}+{v_2+k_2\over v_2-k_2}\,{\tilde\zeta_2\over k_1}
+{1\over k_1k_2}{(v_1-v_2)^2-(k_1-k_2)^2\over (v_1-v_2)^2-(k_1+k_2)^2}, \eqno(3.15b)$$
$$g_0=|g_0|\,e^{{\rm i}\chi},\quad |g_0|^2={(v_1-k_1)(v_2-k_2)\over(v_1+k_1)(v_2+k_2)}. \eqno(3.15c)$$

\begin{figure}[h]
\begin{minipage}{0.22\textwidth}
\includegraphics[width=\textwidth]{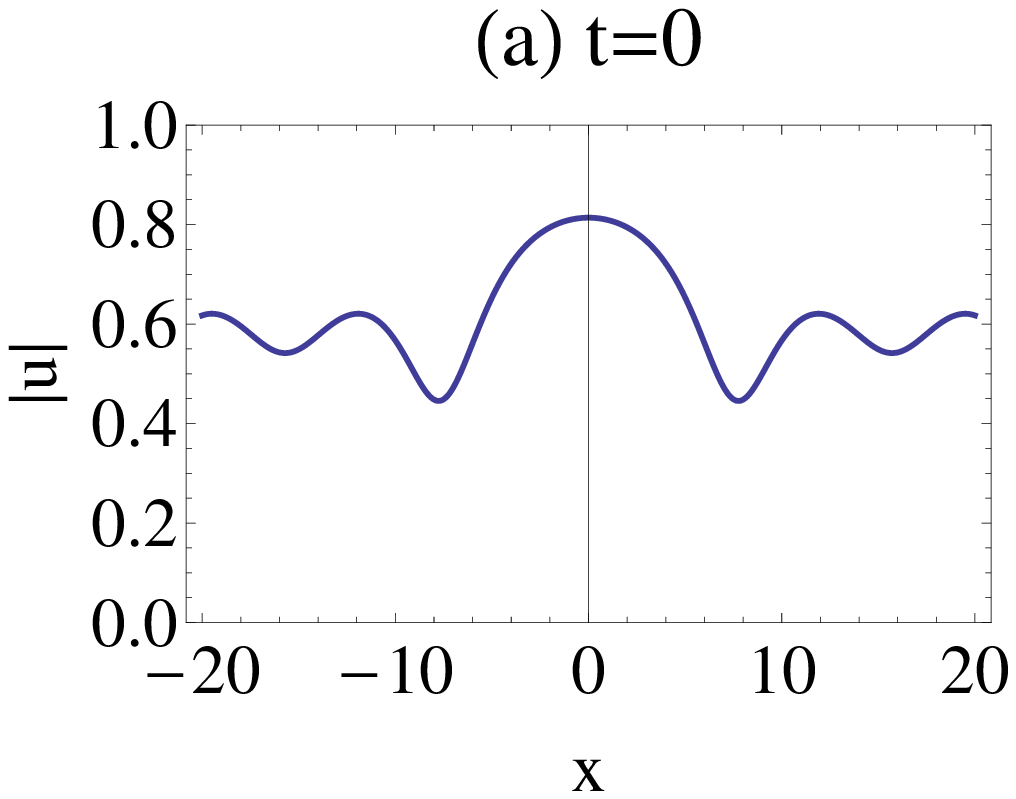}
\end{minipage}
\hspace{9.6pt}
\begin{minipage}{0.22\textwidth}
\includegraphics[width=\textwidth]{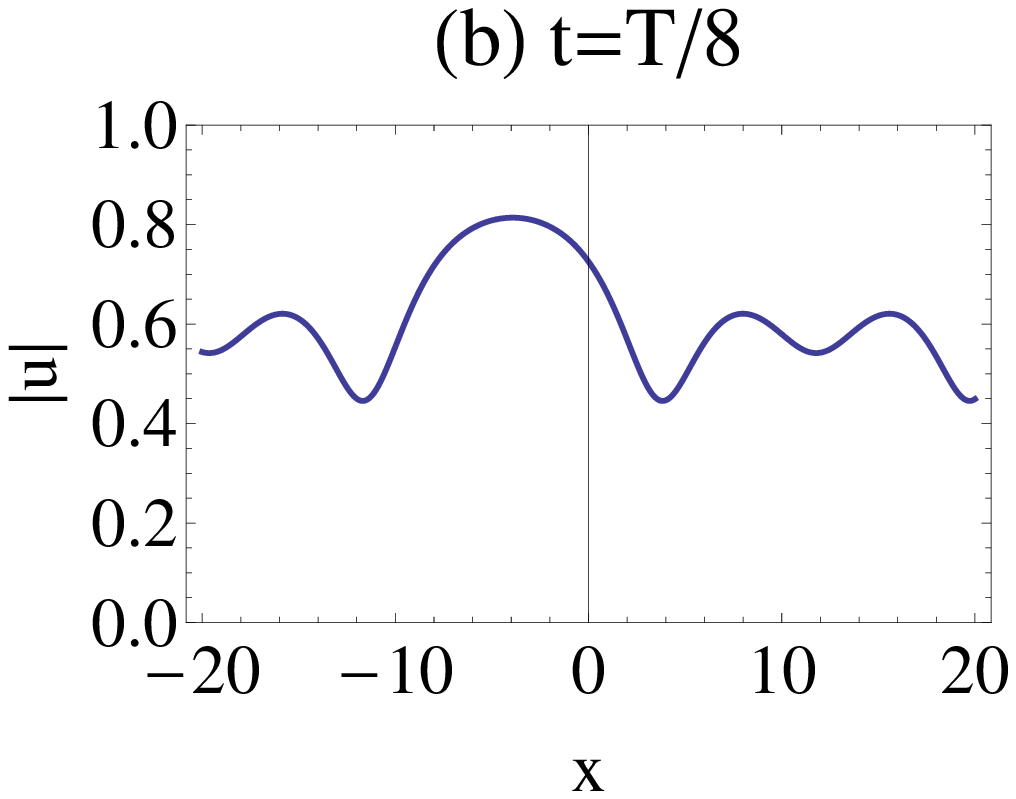}
\end{minipage}
\hspace{9.6pt}
\begin{minipage}{0.22\textwidth}
\includegraphics[width=\textwidth]{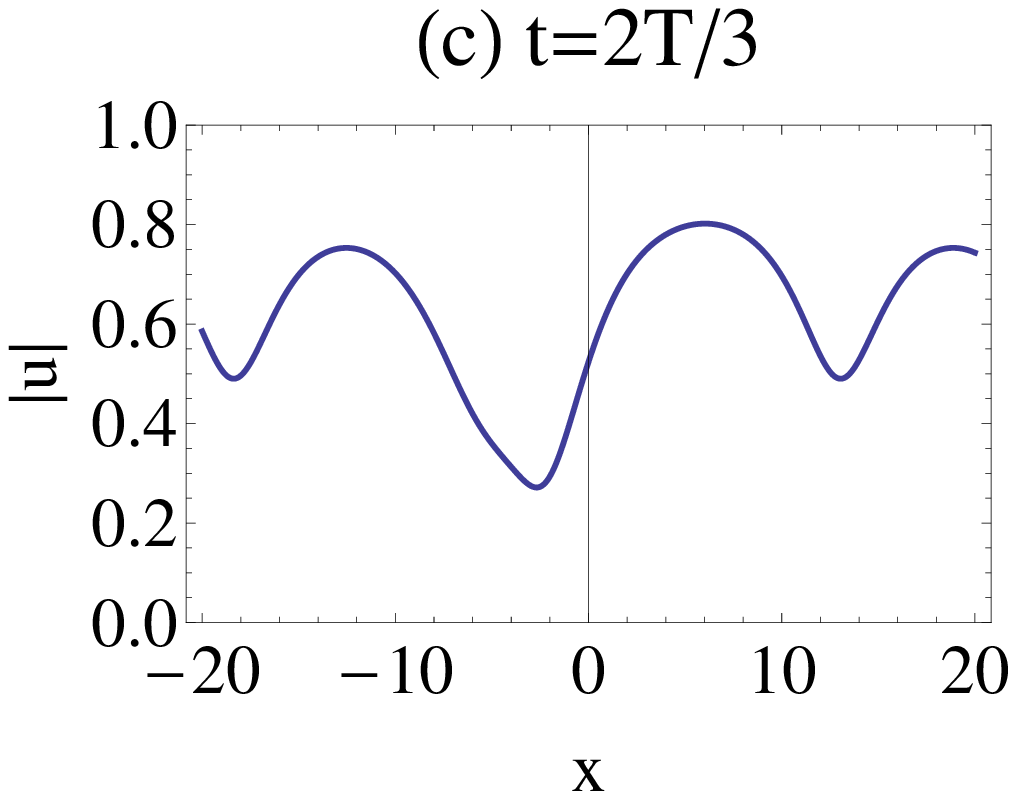}
\end{minipage}
\hspace{9.6pt}
\begin{minipage}{0.22\textwidth}
\includegraphics[width=\textwidth]{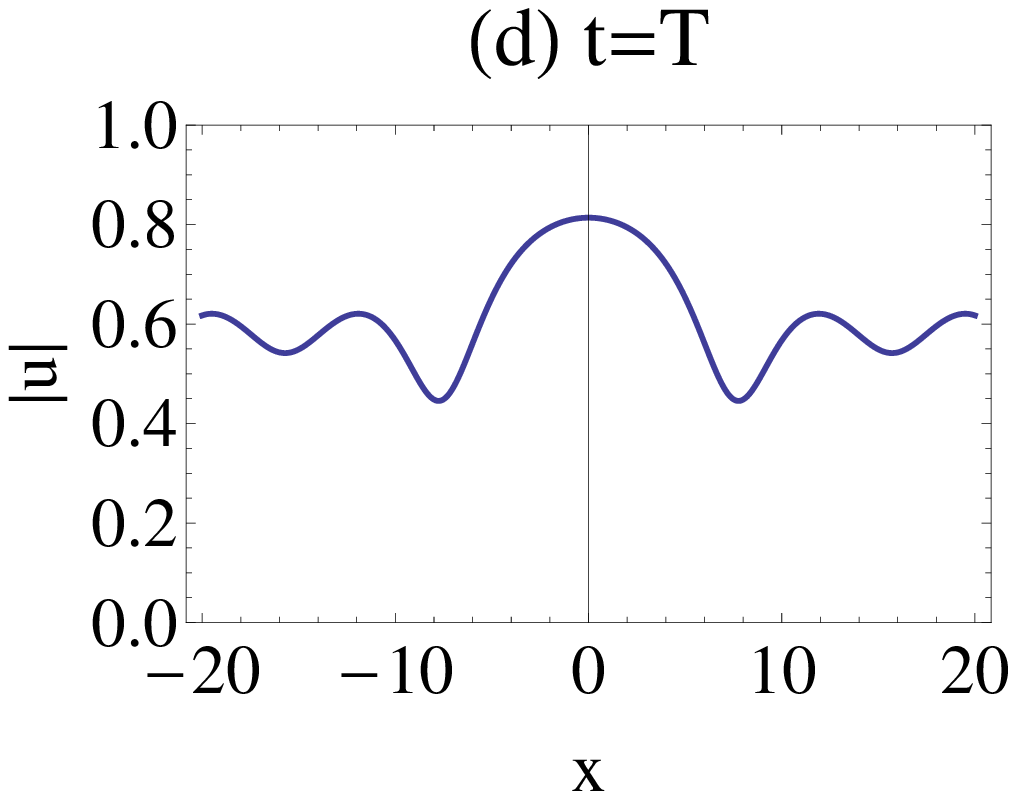}
\end{minipage}
\noindent  {\bf Figure 3.}\ The profiles of the modulus  $|u|$ are plotted for the two-phase solution (3.14b) as function of $x$.
The parameters are set to $ k_1=0.2, k_2=0.4, v_1=-0.3, v_2=-1.5, x_{10}=x_{20}=0$. The period  is $T=100\pi/3$.
 The recurrence time is $T_R=T/8$.\par
\end{figure}
Figure 3 illustrates the profiles of $|u|$ from (3.14b). 
In this example, $T_1=100\pi/3, T_2=10\pi/3$ and $T_R=25\pi/6$.   The period $T$ turns out to be $T_1$.
\par
\bigskip
\leftline{\bf 4. Reductions of the $N$-phase solution} \par
\medskip
\noindent Here, we present some results arising from the reductions of the $N$-phase solution constructed in section 2. 
First, we consider the special case of the solution in which  all the wavenumbers coincide. The resulting solution is shown to exhibit a
pole representation whose pole dynamics obey an integrable Calogero-Moser-Sutherland dynamical system. The second reduction is concerned with
the rational (or algebraic) $N$-soliton solution. It is demonstrated that  the $N$-soliton solution can be reduced simply from the $N$-phase solution 
via an appropriate limiting procedure. 
\par
\medskip
\noindent {4.1. Special case of the $N$-phase solution} \par
\medskip
\noindent The $N$-phase solution takes a simple structure when all the wavenumbers $k_j\ (j=1, 2, ..., N)$ are equal to a positive constant $k$, for example.
 In this setting, the tau-function $f$ takes the form
$$f=|F|,\quad F=(f_{jl})_{1\leq j,l\leq N}, \quad f_{jl}=\zeta_j\delta_{jl}+{2\over v_j-v_l+2k}, \eqno(4.1a)$$
$$\zeta_j={e^{-{\rm i}\theta_j+\delta_j}\over k}={e^{-{\rm i}k\xi_j+\delta_j}\over k}, \quad \xi_j=x-v_jt-x_{j0}, \quad (j=1, 2, ..., N), \eqno(4.1b)$$
$$e^{A_{jl}}={(v_j-v_l)^2\over (v_j-v_l)^2-4k^2}, \quad (j, l=1, 2, ..., N; j\not=l), \eqno(4.1c)$$
whereas the tau-function $g$ is constructed simply from $f$ in accordance with a prescription given in (2.2).
It turns out that the solution $u$ can be represented by a rational function of $e^{{\rm i}kx}$. In particular, the tau-function $f$ becomes an $N$th-order polynomial of $e^{{\rm i}kx}$.
This fact makes it possible to introduce an ansatz for the solution.
The proposition below characterizes the structure of a special class of the $N$-phase solution. \par
\bigskip
\noindent {\bf Proposition 4.1.}\ {\it The periodic solution (2.1)  with (4.1) admits a pole representation
$$u=\beta +{\rm i}\, \sum_{j=1}^Nc_j\Phi(x-x_j),\quad \Phi(x)={k\over 2}\,\cot\,{kx\over 2},\quad \beta \in\mathbb{R}, \quad {\rm Im}\,x_j<0,\quad (j=1, 2, ..., N). \eqno(4.2)$$
  The pole $x_j=x_j(t)$ and the residue $c_j=c_j(t)$ evolve according to the system of nonlinear
ordinary differential equations (ODEs)
$$\dot x_j={2\beta\over c_j}+{2{\rm i}\over c_j}\, \sum_{\substack{k=1\\ (k\not=j)}}^Nc_k\Phi'(x_j-x_k),\quad (j=1, 2, ..., N), \eqno(4.3)$$
$$\dot c_j=-2{\rm i}\,\sum_{\substack{k=1\\ (k\not=j)}}^Nc_k^*\Phi'(x_j-x_k), \quad (j=1, 2, ..., N), \eqno(4.4)$$
where the dot denotes the differentiation with respect to $t$ and $\Phi'(x)=d\Phi(x)/dx$.
In addition, the $N$ constraints
$$1-\beta c_j+{\rm i}c_j\,\sum_{k=1}^N\Phi(x_j-x_k^*)c_k^*=0,\quad (j=1, 2, ..., N), \eqno(4.5)$$
are imposed on $c_j$ and $x_j\ (j=1, 2, ..., N)$.}
\par
\medskip
\noindent {\bf Proof.}\  Substituting (4.2) into equation (1.1) and comparing the coefficients of $\Phi'(x-x_j), \Phi(x-x_j)$ and $\Phi^3(x-x_j)$ at the pole $x=x_j$, we obtain
(4.3), (4.4) and (4.5), respectively.  \hspace{\fill}$\Box$ 
\par
\medskip
\noindent {\bf Remark 4.1.}\ The similar system has been derived for the periodic solutions of equation (1.2) [5].  It is important
that the constraints (4.5) are preserved if they hold at an initial time. With the aid of (4.5), one can  eliminate the variables $c_j$ from (4.3) and (4.4).  The resulting  system of second order ODEs for
$x_j$ turns out to be the completely integrable Calogero-Moser-Sutherland dynamical system [20-25]
$${d^2x_j\over dt^2}=-{\partial\over \partial x_j}\,\sum_{\substack{k=1\\ (k\not=j)}}^N{k^2\over \sin^2\left[{k\over 2}(x_j-x_k)\right]}, \quad (j=1, 2, ..., N). \eqno(4.6)$$
An exact method of solution developed in [5] for solving an analogous system of equations associated with  equation (1.2) can be applicable as well to the above system of equations.
In this approach, the major nontrivial issue to be addressed is to establish that the poles $x_j(t)$ stay in the lower-half complex plane as time evolves.
Fortunately, this problem has been resolved in section 2  by constructing explicitly an $N$-phase solution together with its analyticity in the upper-half complex plane.
\par
\bigskip
\noindent {4.2. $N$-soliton solution} \par
\medskip
\noindent The $N$-soliton solution is produced in the long-wave limit of the $N$-phase solution, as we now demonstrate. \par
\medskip
\noindent {\bf Proposition 4.2.}\ {\it The $N$-soliton solution of equation (1.1) can be represented by the determinantal formulas
$$u= {\tilde g\over \tilde f},\quad  |u|^2=-{\rm i}\,{\partial\over \partial x}\,{\ln}\,{\tilde f^*\over \tilde f}, \quad \tilde f=|\tilde F|,\quad \tilde g=\tilde g_0|\tilde G|, \eqno(4.7a)$$
where 
$$ \tilde F=(\tilde f_{jk})_{1\leq j,k\leq N}, \quad \tilde f_{jk}=(\xi_j+{\rm i}\rho\delta_{j1})\delta_{jk}-{2{\rm i}\over v_j-v_k}(1-\delta_{jk}), \eqno(4.7b)$$
$$\tilde G=(\tilde g_{jk})_{1\leq j,k\leq N}, \quad \tilde g_{1k}=1\ (k=1, 2, ..., N), \quad \tilde g_{jk}=\tilde f_{jk}\quad (j=2, 3, ..., N, k=1, 2, ..., N). \eqno(4.7c)$$
$$ \tilde g_0=-{\rm i}\sqrt{2\rho}\,e^{{\rm i}\chi}, \quad \xi_j=x-v_jt-x_{j0}, \ (v_1=0,\  v_j>0,\ j=2, 3, ..., N), \eqno(4.7d)$$
The tau-functions $\tilde f$ and $\tilde g$ satisfy the bilinear equations
$${\rm i}D_t\tilde g\cdot \tilde f=D_x^2\tilde g\cdot \tilde f, \eqno(4.8a)$$
$${\rm i}D_x\tilde f^*\cdot \tilde f=-\tilde g^*\tilde g. \eqno(4.8b)$$ }
\par
\medskip
\noindent {\bf Proof.}\ The $N$-soliton solution is derived simply from  the long-wave limit of the $N$-phase solution. To be more specific, one first shifts the phase variables $x_{j0}$ as
$x_{j0}\rightarrow x_{j0}+\pi/k_j\ (j=1, 2, ..., N)$   and then takes the limits $k_j\rightarrow 0$ while fixing the velocities $v_j\ (j=1, 2, ..., N)$.
The leading-order asymptotics of the parameters given in (3.4) are found to be as
$$\zeta_j \sim -{1\over k_j}+{\rm i}(\xi_j+{\rm i}\rho \delta_{j1}), \quad \delta_j \sim  \sum_{\substack{l=1\\ (l\not=j)}}^N{2k_jk_l\over (v_j-v_l)^2} +k_1\rho\delta_{j1},
 \quad e^{A_{jl}} \sim 1+{4k_jk_l\over (v_j-v_l)^2}. $$
Consequently, the asymptotics of $f_{jl}, |F|$ and $|G|$ from (3.3) become
$$f_{jj}\sim {\rm i}(\xi_j+{\rm i}\rho\delta_{j1}), \quad f_{jl}\sim {2\over v_j-v_l}\quad (j\not=l),  \quad |F| \sim {\rm i}^N|\tilde F|, $$
$$|g_0|\sim \sqrt{2\rho}, \quad  |G|\sim {\rm i}^{N-1}|\tilde G|.$$
Substituting these expressions into (3.2) and taking the limits $k_j\rightarrow 0$, we obtain (4.7a). 
The bilinear equations (4.8) stem simply from (2.6) and (2.7) by introducing the above asymptotic expressions of $|F|$ and $|G|$. \hspace{\fill}$\Box$ \par
\medskip
\noindent {\bf Remark 4.2.}\ The $N$-soliton solution given above recovers the corresponding one obtained in [11] by using
an inverse spectral formula for the Lax operator. Note that since $p_1=0$, $v_1=-k_1\rightarrow 0$ as $ k_1\rightarrow 0$. 
The pole representation of the $N$-soliton solution subjected to the boundary condition $u\rightarrow 0,\ |x|\rightarrow \infty$ follows from
(4.2)-(4.4) by taking the limit $k\rightarrow 0$. It reads in the form
$$u=\sum_{j=1}^N{c_j\over x-x_j}. \eqno(4.9)$$
The time evolution of $x_j$ and $c_j$ is governed by the system of ODEs
$$\dot x_j={2{\rm i}\over c_j}\,\sum_{\substack{k=1\\ (k\not=j)}}^N{c_k\over x_j-x_k}, \quad \dot c_j={2{\rm i}}\,\sum_{\substack{k=1\\ (k\not=j)}}^N{c_j-c_k\over (x_j-x_k)^2},
  \quad (j=1, 2, ..., N), \eqno(4.10)$$
  under the constraints
$$\quad 1+{\rm i}c_j\,\sum_{j=1}^N{c_k^*\over x_j-x_k^*}=0,\quad (j=1, 2, ..., N). \eqno(4.11)$$
When plugged into (4.11), the system of ODEs (4.10) for $x_j$ can be recast to the Calogero-Moser system [20, 22, 23]
$${d^2x_j\over dt^2}=-4\,{\partial\over \partial x_j}\,\sum_{\substack{k=1\\ (k\not=j)}}^N{1\over (x_j-x_k)^2}, \quad (j=1, 2, ..., N). \eqno(4.12)$$
\par
\medskip
\noindent {\bf Remark 4.3.}\ The long-wave limit of the $N$-phase solution (3.4) with (3.5) for equation (1.2) takes the form [3, 9]
$$u= {\tilde g\over \tilde f},\quad  |u|^2=1+{\rm i}\,{\partial\over \partial x}\,{\ln}\,{\tilde f^*\over \tilde f},\quad \tilde f=|\tilde F|, \quad g=\tilde g_0|\tilde G|, \eqno(4.13a)$$
where
$$\tilde F=(\tilde f_{jk})_{1\leq j,k\leq N}, \quad \tilde f_{jk}=\left(\xi_j+{{\rm i}\over a_j}\right)\delta_{jk}-{2{\rm i}\over v_j-v_k}(1-\delta_{jk}), \eqno(4.13b)$$
$$\tilde G=(\tilde g_{jk})_{1\leq j,k\leq N}, \quad \tilde g_{jk}=\left\{\xi_j+{\rm i}\left({1\over a_j}+{2\over v_j}\right)\right\}\delta_{jk}-{2{\rm i}\over v_j-v_k}(1-\delta_{jk}), \eqno(4.13c) $$
$$\tilde g_0=e^{{\rm i}\chi}, \quad \xi_j=x-v_jt-x_{j0},  \quad (j=1, 2, ..., N). \eqno(4.13d)$$
The amplitude parameters $a_j$ are related to the velocity parameters $v_j$ by the relations
$$v_j(v_j+2)+2a_j=0, \quad 0<a_j<{1\over 2}, \quad \quad (j=1, 2, ..., N). \eqno(4.14)$$
\bigskip
\noindent {4.3. Examples of solutions}\par
\medskip
\noindent We explore  the properties of soliton solutions in comparison with those of equation (1.2). 
Specifically, we consider the one- and two-soliton solutions. \par
\medskip 
\noindent {4.3.1. One-soliton solution}\par
\bigskip
\noindent The one-soliton solution of equation (1.1) takes the form
$$u={-{\rm i}\sqrt{2\rho}\,e^{{\rm i}\chi}\over x-x_{10}+{\rm i}\rho}, \eqno(4.15a)$$
$$|u|^2={2\rho\over (x-x_{10})^2+\rho^2}, \eqno(4.15b)$$
and the corresponding solution of equation (1.2) is written in the form
$$u={e^{{\rm i}\chi}\over x-v_1t-x_{10}+{{\rm i}\over a_1}}, \eqno(4.16a)$$
$$|u|^2=1-{2a_1\over a_1^2(x-v_1t-x_{10})^2+1}, \eqno(4.16b)$$
where $v_1(v_1+2)+2a_1=0$ and $0<a_1<1/2$. \par
\begin{center}
\includegraphics[width=8cm]{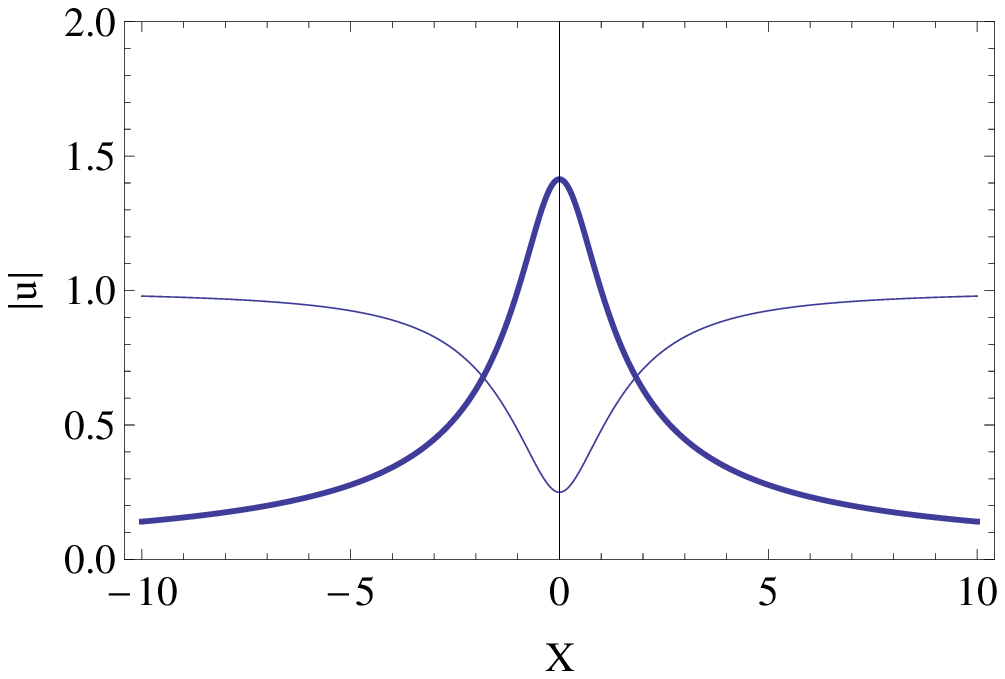}
\end{center}
\noindent  {\bf Figure 4.}\ The long-wave limit of the one-phase solutions (3.9) (thick curve) and (3.10) (thin curve) as function of the traveling wave coordinate $X=x-v_1t-x_{10}$.  
The   parameters are same as those used in  figure 1. Note from (4.14) with $v_1=-3/4$ that $a_1=15/32$. \par
\bigskip
Figure 4 depicts the long-wave limit of one-phase solutions   which are given respectively by
(4.15a) and (4.16a).
 The former solution takes the form of a bright soliton whereas the latter one shows a dark soliton on a constant background. 
 It is interesting to observe that the velocity of the bright soliton becomes zero since $v_1\rightarrow 0$ in the long-wave limit.
 This reflects the fact that the eigenvalue problem (1.3) for the $N$-soliton potential has zero eigenvalue [11].  Note that in view of 
 the invariance of equation (1.1) under the Galilean transformation, 
 $u(x, t)\rightarrow e^{{\rm i}vx-{\rm i}v^2t}\,u(x-2vt, t)$, one can rewrite (4.15a) in a time-dependent form presented in [10].
  \par
 \medskip
 \noindent {4.3.2. Two-soliton solution}\par
 \bigskip
 \begin{figure}[h]
\begin{minipage}{0.22\textwidth}
\includegraphics[width=\textwidth]{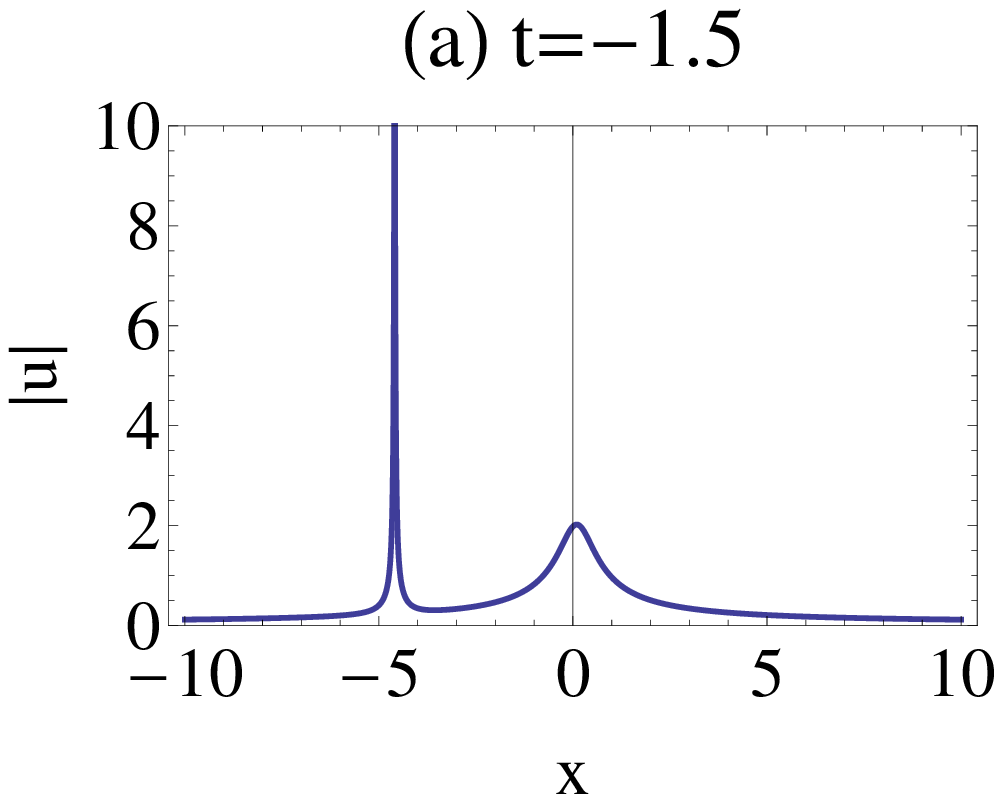}
\end{minipage}
\hspace{9.6pt}
\begin{minipage}{0.22\textwidth}
\includegraphics[width=\textwidth]{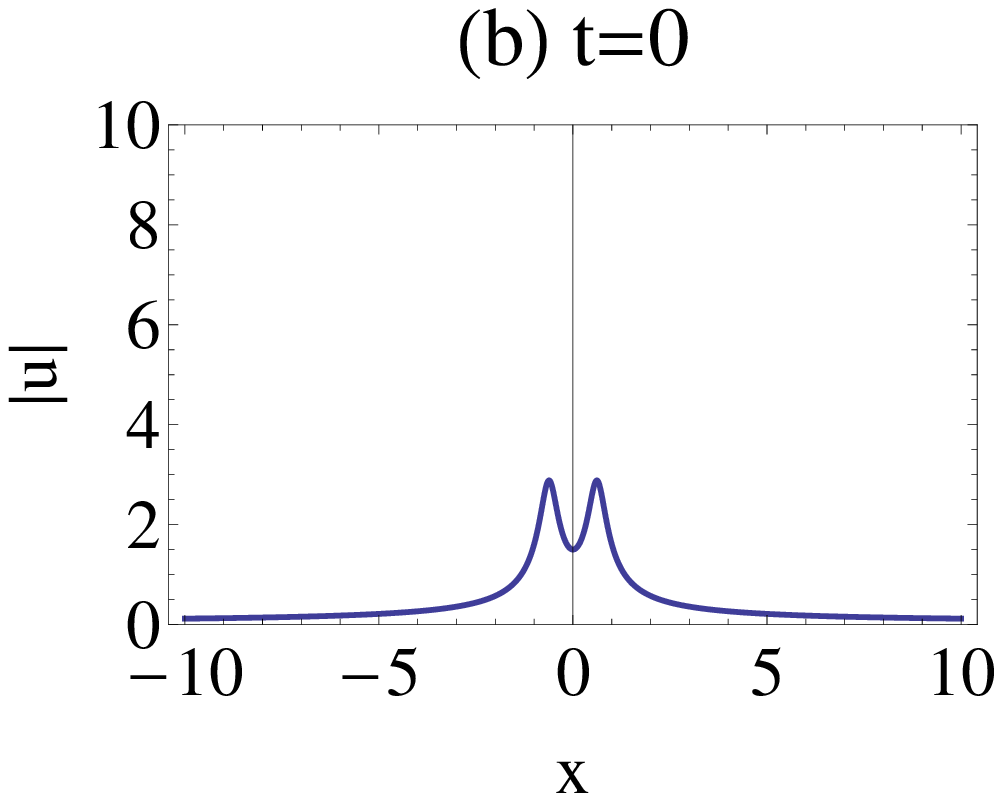}
\end{minipage}
\hspace{9.6pt}
\begin{minipage}{0.22\textwidth}
\includegraphics[width=\textwidth]{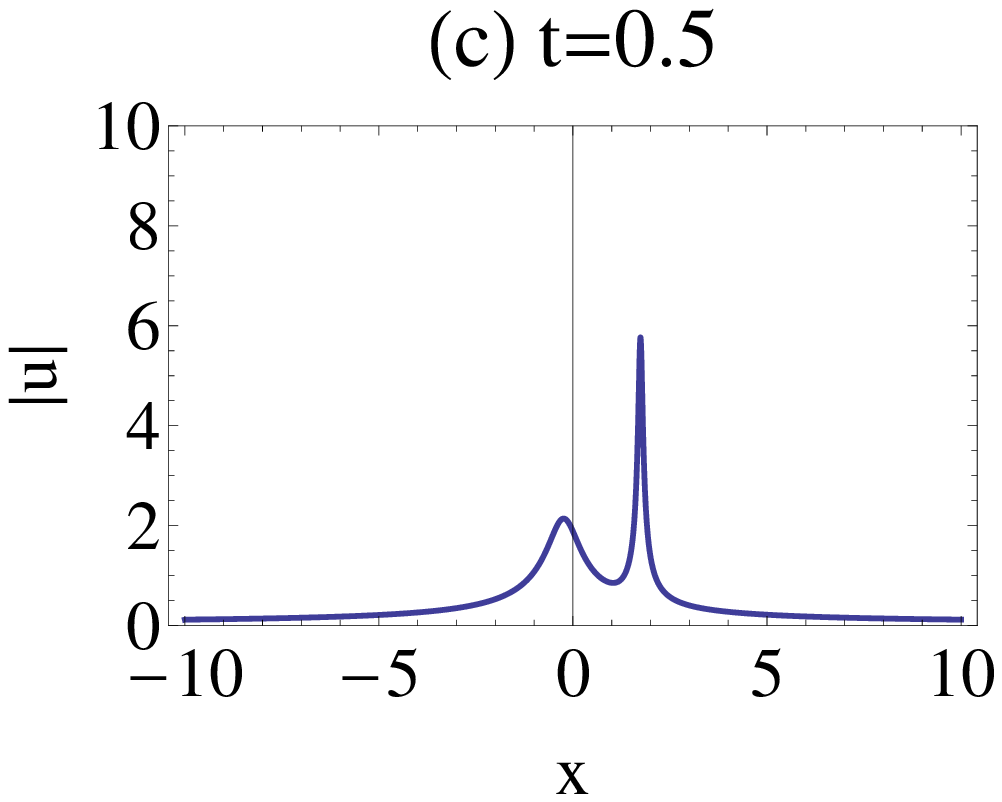}
\end{minipage}
\hspace{9.6pt}
\begin{minipage}{0.22\textwidth}
\includegraphics[width=\textwidth]{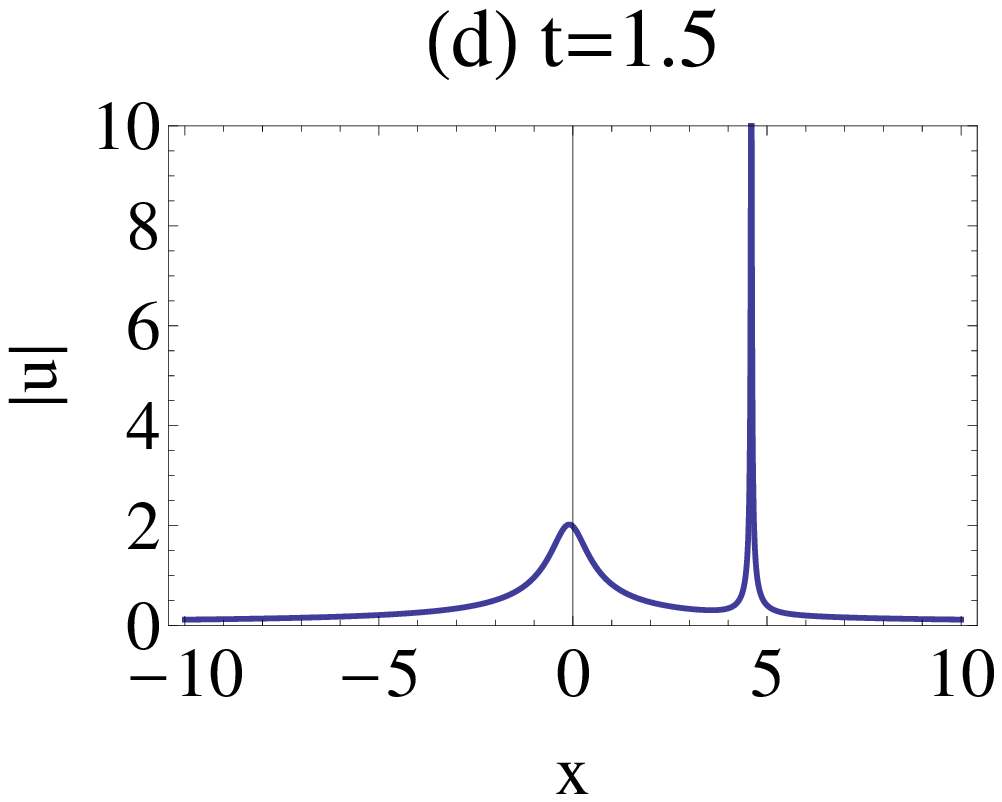}
\end{minipage}
\par
\noindent {\bf Figure 5.}\ The two-soliton solution of equation (1.1)  as function of $x$.  
The   parameters are the  same as those used in  figure 2 except $v_1=0$.\par
\end{figure}
\par
\begin{figure}[h]
\begin{minipage}{0.22\textwidth}
\includegraphics[width=\textwidth]{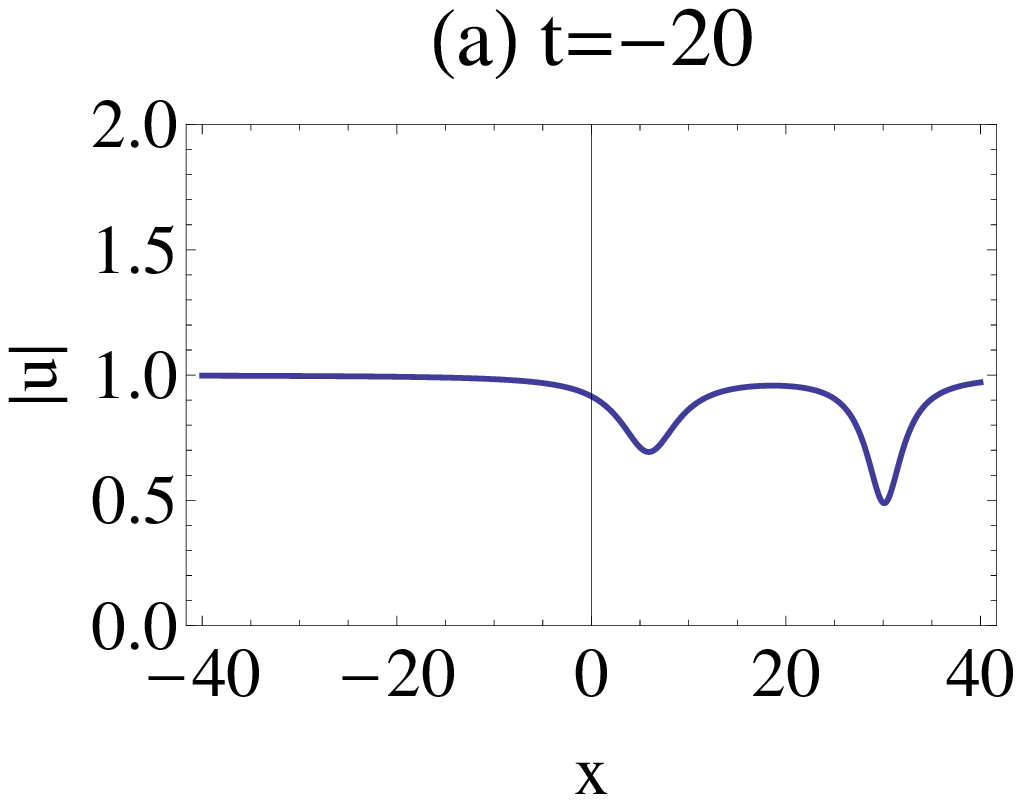}
\end{minipage}
\hspace{9.6pt}
\begin{minipage}{0.22\textwidth}
\includegraphics[width=\textwidth]{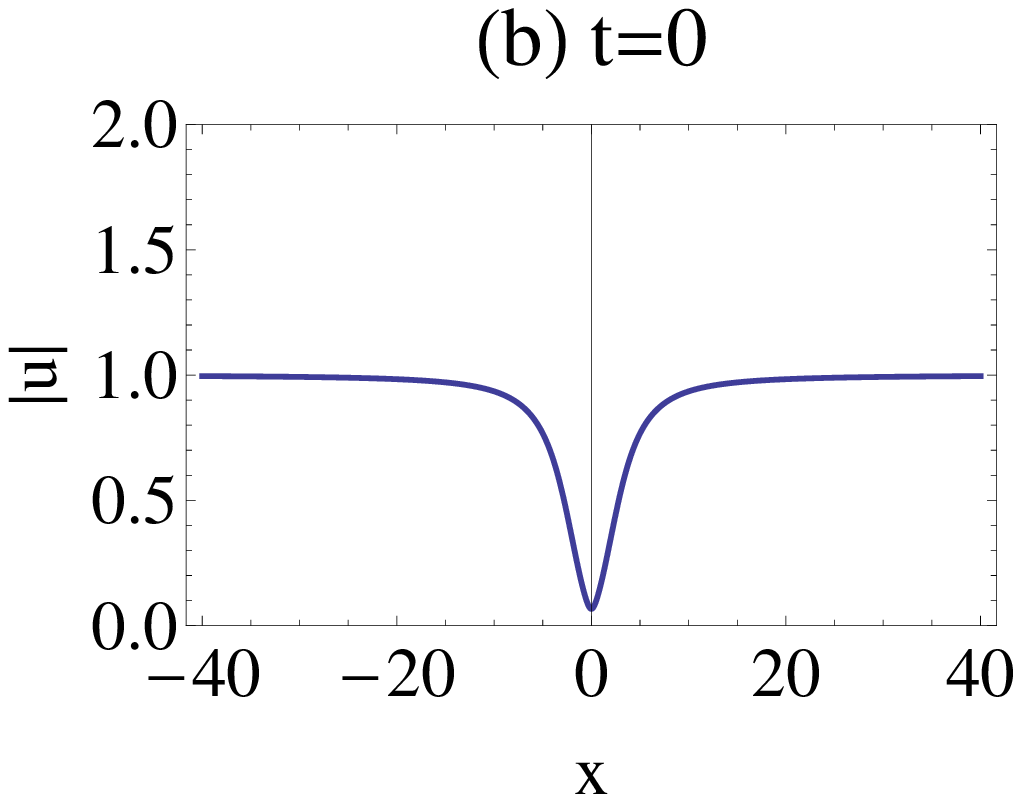}
\end{minipage}
\hspace{9.6pt}
\begin{minipage}{0.22\textwidth}
\includegraphics[width=\textwidth]{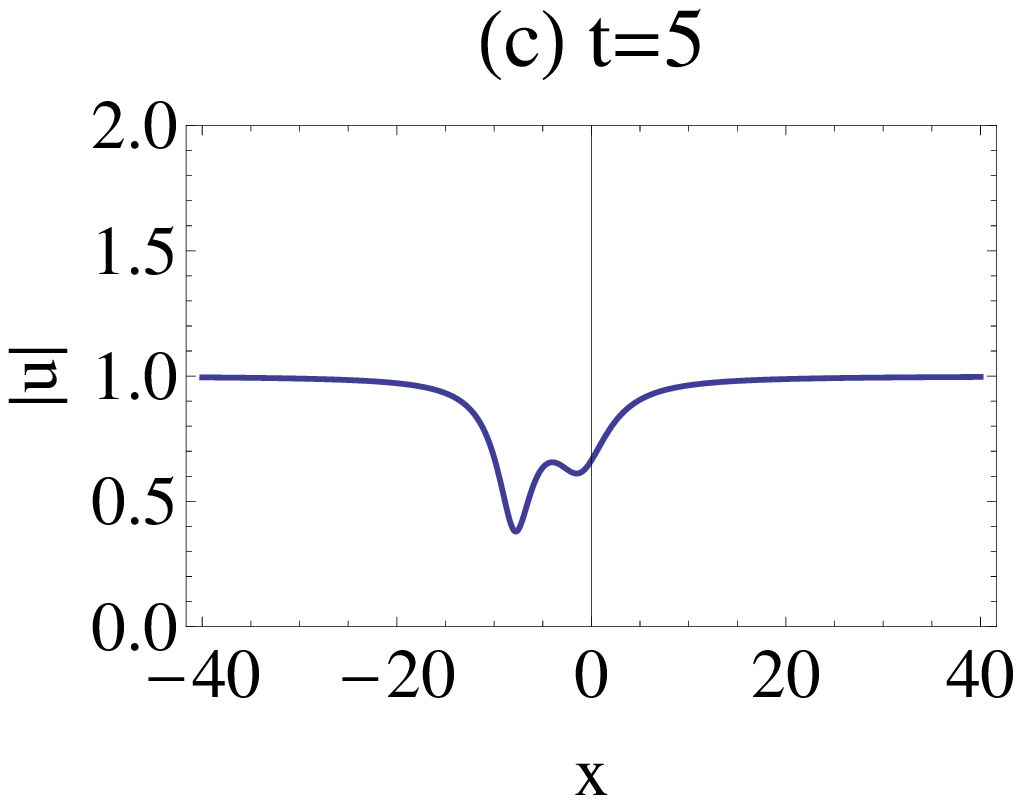}
\end{minipage}
\hspace{9.6pt}
\begin{minipage}{0.22\textwidth}
\includegraphics[width=\textwidth]{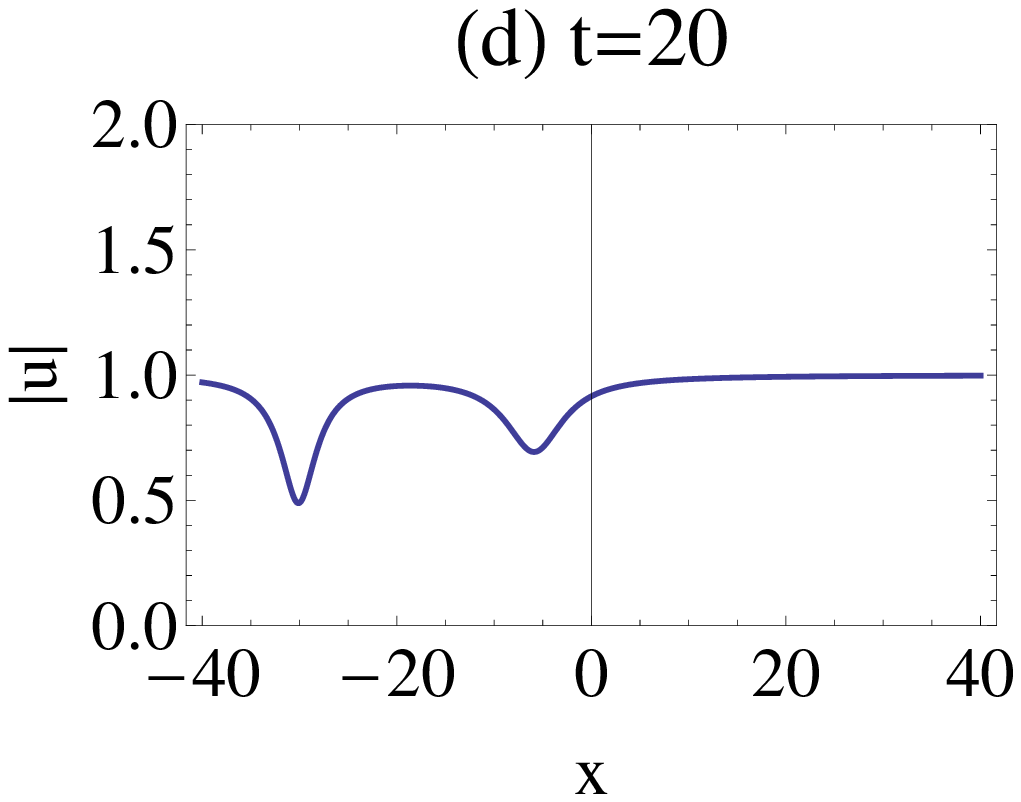}
\end{minipage}
\par
\noindent  {\bf Figure 6.}\  The two-soliton solution of equation (1.2) as function of $x$.  
The  parameters are the same as those used in  figure 3.\par
\end{figure}
\bigskip
 \noindent Figure 5 shows the two-soliton solution of equation (1.1) which has been  obtained by the long-wave limit of the
 two-phase solution depicted in figure 2. See (4.7). It represents the interaction of two solitons. The asymptotic profile of the solution consists of a superposition of
 the two solitons, one is a bright soliton with zero velocity and the other one takes the form of the  delta function. Explicitly,
 $$|u|^2 \sim {2\rho\over (x-x_{10})^2+\rho^2} +2\pi\delta(x-v_2t-x_{20}), \quad t \rightarrow \pm\infty. \eqno(4.17)$$
 We recall that this intriguing feature of the solution has been found in analyzing the structure of the two-soliton solution of equation (1.1) [11]. \par
 Figure 6 shows the two-soliton solution of equation (1.2) reduced from  the long-wave limit of the two-phase solution depicted in figure 3. See (4.13).
 The asymptotic profile is simply a superposition of two dark solitons. It reads
  $$|u|^2 \sim 1- \sum_{j=1}^2{2a_j\over a_j^2(x-v_jt-x_{j0})^2+1}, \quad t \rightarrow \pm\infty. \eqno(4.18)$$
  In the above two examples, solitons exhibit no phase shifts after their interaction. The similar feature 
  has been found for the first time in  the interaction process of rational  solitons of the BO equation [26, 27]. \par
  \bigskip
 \noindent {\bf Remark 4.4.}\  An explicit expression of the $N$-soliton solution of equation (1.1) has been derived in
  the analysis of the spectral problems of the Lax operator [11]. 
  The detailed investigation of the dynamics of solitons was performed as well.
   In particular,  the asymptotic form of the $N$-soliton solution at large time was shown to be represented by a superposition of a single bright  soliton with zero velocity
   and a train of $N-1$ pulses with delta function profiles. To be more specific, 
   $$|u|^2  \sim {2\rho\over (x-x_{10})^2+\rho^2}+2\pi\sum_{j=2}^N\,\delta(x-v_jt-x_{j0}), \quad t \rightarrow \pm\infty. \eqno(4.19)$$
  Obviously, this expression is a generalization of the asymptotic of the two-soliton solution  (4.17). \par
  We recall that the similar  behavior of solution has been observed in the interaction process of solitons of the following sine-Hilbert equation [28-33]
    $$H\theta_t = -\sin\,\theta, \quad \theta=\theta(x, t). \eqno(4.20)$$
  The asymptotic form of the $N$-soliton solution has been found to be as
    $$\theta_x \sim {2a_1\over (x-a_1t-b_1+a_2/a_1)^2+a_1^2}+2\pi\sum_{j=1}^{N-1}\,\delta(x-\alpha_j), \quad t\rightarrow \pm\infty, \eqno(4.21)$$
  where $a_1>0, a_2, b_1, \alpha_j (j=1, 2, ..., N-1) \in \mathbb{R}$. Unlike the asymptotic form (4.19), a single soliton propagates with a constant velocity 
  and remaining $N-1$ pulses take the form of the delta functions with zero velocity. It is interesting 
  to see that the velocity of the soliton is inversely proportional
  to its amplitude so that the small soliton propagates more rapidly than the large soliton.
  \par
  \bigskip
  \noindent {4.4. Alternative representation of the $N$-soliton solution}\par
\medskip
\noindent The $N$-soliton solution presented in proposition 4.2 has an alternative representation
  which is given in  the proposition. \par
\medskip
\noindent {\bf Proposition 4.3.}\ {\it The squared modules of the $N$-soliton solution of equation (1.1) admits a representation
$$|u|^2=\sum_{j=1}^Nv_j-{\rm i}\,\sum_{j=1}^N(\tilde\mu_j-\tilde\mu_j^*), \eqno(4.22)$$
where the functions $\tilde\mu_j=\tilde\mu_j(x, t)$ solve the system of nonlinear algebraic equations
$$\sum_{k=1}^N{2{\rm i}\over v_j-2{\rm i}\tilde\mu_k}= \xi_j+{\rm i}\rho\delta_{j1}+\sum_{\substack{k=1 \\ (k\not= j)}}^N {2{\rm i}\over v_j-v_k}, \quad (j=1, 2, ..., N). \eqno(4.23)$$
They satisfy the system of nonlinear PDEs
$$\sum_{k=1}^N{\tilde\mu_{k,x}\over (v_j-2{\rm i}\tilde\mu_k)^2}=-{1\over 4}, \quad \sum_{k=1}^N{\tilde\mu_{k,t}\over (v_j-2{\rm i}\tilde\mu_k)^2}={v_j\over 4},\quad (j=1, 2, ..., N). \eqno(4.24)$$}
\par
\medskip
\noindent  {\bf Proof.}\ If we apply the long-wave limit prescribed in proposition 4.2 to proposition 2.3, the above results follow immediately.  \hspace{\fill}$\Box$ \par
\medskip
\noindent{4.5. Eigenvalue problem for the $N$-soliton solution} \par
\medskip
\noindent The eigenvalue problem  for the $N$-soliton solution has been resolved by the spectral analysis of the 
Lax operator associated with equation (1.1). For completeness, we reproduce the result by means of the reduction procedure.
\par
\medskip
\noindent {\bf Proposition 4.4.} {\it Let $\phi_ j, \psi_j^+$  and $\lambda_j$ be the eigenfunctions and corresponding eigenvalue
for equation (1.3
) with the $N$-soliton solution (4.7). If  $\phi_ j$ and $ \psi_j^+$  satisfy the system of linear algebraic equations
$$\sum_{k=1}^N\tilde f_{jk}\phi_k=g_0\delta_{j1}, \quad (j=1, 2, ..., N), \eqno(4.25a) $$
$$ \sum_{k=1}^N\tilde f_{jk}\psi_k^+={\rm i}, \quad (j=1, 2, ..., N), \eqno(4.25b)$$
then they solve the eigenvalue equations 
$${\rm i}\phi_{j,x}+\lambda_j\phi_j+u\psi_j^+=0, \quad (j=1, 2, ..., N),\eqno(4.26)$$
$${\rm i}\phi_{j,t}-2{\rm i}\lambda_j \phi_{j,x}+\phi_{j,xx}-2{\rm i}u_x\psi_j^+=0, \quad (j=1, 2, ..., N),\eqno(4.27)$$
$${\rm i}\psi^+_{j,t}-2{\rm i}\lambda_j \psi^+_{j,x}+\psi^+_{j,xx}-{\rm i}[(1-{\rm i}H)(|u|^2)_x]\psi_j^+=0, \quad (j=1, 2, ..., N),\eqno(4.28)$$
with $\lambda_j=-{v_j\over 2}\ (j=1, 2, ..., N)$.}
 \par
\medskip
\noindent {\bf Proof.}\ If we take the long-wave limit of (2.32a, b) by employing the leading-order asymptotics of the elements $f_{jk}$ 
given in the proof of proposition 4.2, then
(4.25a, b) follow immediately. Note in this limit that $\lambda_j=-q_j=-(v_j-k_j)/2\rightarrow -v_j/2$, as indicated. 
The proof of (4.26)-(4.28) can be done in parallel with the proof of equations (2.33)-(2.35).  
Actually, after performing the limiting procedure, it is found that they take the same forms as (2.33)-(2.35) .
\hspace{\fill}$\Box$ \par 
\bigskip
\noindent {\bf Remark 4.5.}\ An intriguing result extracted from proposition 4.4 is that $v_1=0$, as a consequence of the relations $v_1=p_1+k_1, p_1=k_1=0$.
Thus, among $N$ solitons, one soliton does not propagate. One can observe this phenomenon in the interaction process of two-solitons.
See figure 5. The eigenfunction $\phi_1$ for the eigenvalue $\lambda_1=0$ takes the form $\phi_1=c(\tilde f-\tilde f^*)/\tilde f$, where  $c\in\mathbb{C}$
is a normalization constant. Indeed, it follows from
(1.3) and (1.4) that $\phi_1$ satisfies the equation  ${\rm i}\phi_{1,x}+uP_+(u^*\phi_1)=0$ with $P_+={1\over 2}(1-{\rm i}H)$ being a projection operator.
 One can show that if  $\phi_1$ and $u^*$ from (4.7a) are substituted into the undermentioned 
equation for $\phi_1$, then it recasts to the bilinear equation (4.8b). \par
\bigskip
\noindent {\bf 5. Concluding remarks} \par
  \medskip
  \noindent In this paper, we were concerned with the focusing nonlocal NLS equation. Specifically, we  constructed the $N$-phase solution by means of the
  direct method. The long-wave limit   was then  taken for the $N$-phase solution  to deduce  the $N$-soliton solution. 
  We illustrated  a few examples of both the periodic and soliton solutions in comparison with those corresponding to the  defocusing nonlocal NLS equation
  and clarified their novel features.  
  \par
 There are still many problems we have to settle, some of them are listed below: \par
 \medskip
 \noindent 1. The initial value problem will be the most important issue which may be tackled with the aid of IST.
  As a first step toward this goal,  the eigenfunctions of  the linear system (1.3)-(1.6) were obtained  for both the $N$-phase and $N$-soliton potentials.
  Recall that the method has been applied to the defocusing nonlocal NLS equation to solve the direct and inverse scattering problems 
    for the $N$-soliton potential [7, 8]. \par
    \medskip
 \noindent 2. Equation (1.2) has been derived formally by using an asymptotic multiscale expansion method to the BO equation [2].
    It is interesting to explore whether a similar procedure is applicable for deriving equation (1.1) as well starting from an integrable nonlocal evolution equation.     
\par
\medskip
\noindent 3. There are several ways to extend the nonlocal NLS equations (1.1) and (1.2). 
On direction is the extension to the finite depth case. Actually, their exists a finite depth analog of (1.2) in which the 
Hilbert transform is replaced by the operator $T$ defined by $Tu(x, t)={1\over 2\delta}P\int_{-\infty}^\infty\coth\left[{\pi(y-x)\over 2\delta}\right]u(y, t)dy$, where
$\delta$ is a parameter characterizing the depth of fluid [34]. The analysis of the resulting equation by IST has been developed, showing its complete integrability.
Furthermore, the $N$-soliton solution has been derived by a direct method [35]. In regards to (1.2), on the other hand, its formal extension has been done in [10] but its 
detailed analysis still remains open except for the construction of an $N$-soliton solution [10]. \par
\medskip
\noindent 4. The extension to the multi-component systems is also worth studying. As was mentioned briefly in introduction,  the intermediate mixed Manakov system  combines the
finite depth versions of (1.1) and (1.2) [15]. By introducing an ansatz, this system is shown to be connected to the integrable Calogero-Moser dynamical system with the
hyperbolic and elliptic potentials. Nevertheless, its complete integrability has not been established yet.

\newpage
\noindent{\bf Appendix A. Integrability and conservation laws}\par
\bigskip
\noindent Here, we first show that equation (1.1) can be obtained as the compatibility conditions of
the system of linear PDEs (1.3)-(1.6).  Then, we derive an infinite number of conservation laws
by solving successively the recursion relation for the eigenfunctions of the linear system. \par
\medskip
\noindent {\bf A.1. Integrability}\par
\medskip
\noindent We differentiate (1.3) by $t$ and use (1.5)  and (1.6) to eliminate $\phi_t, \phi_{xt}$  and $\psi_t^+$, respectively.
Rearranging the resultant expression gives
$$-\phi_{xxx}+3{\rm i}\lambda\phi_{xx}+(2\lambda^2+\kappa)\phi_x-{\rm i}\kappa\lambda \phi
+{\rm i}u\psi_{xx}^++2{\rm i}(u_x\psi^+)_x+ 2\lambda(u\psi_+)_x$$
$$+u\left\{(1-{\rm i}H)(|u|^2)_x-{\rm i}\kappa\right\}\psi^+ +u_t\psi^+=0. \eqno(A.1)$$
It follows by using (1.3) repeatedly that
$$-\phi_{xxx}+3{\rm i}\lambda\phi_{xx}+(2\lambda^2+\kappa)\phi_x-{\rm i}\kappa\lambda \phi=-{\rm i}(u\psi^+)_{xx}-2\lambda(u\psi^+)_x+{\rm i}\kappa u\psi^+. \eqno(A.2)$$
If we substitute  (A.2) into (A.1), we have
$$\left\{u_t+{\rm i}u_{xx}+u(1-{\rm i}H)(|u|^2)_x)\right\}\psi^+=0,$$
which, divided by $\psi^+$, yields equation (1.1). 
We note that  equation (1.4) and the time evolution of $\psi^-$ from (1.6) have not been employed in the derivation process. \par
In a similar manner, we can  derive the evolution  equation for $u^*$. Actually, differentiating (1.4) by $t$ and using (1.5) and (1.6) to eliminate $\phi_t$ and $\psi_t^{\pm}$, we deduce
 $${\rm i}(\psi_{xx}^+ -\sigma \psi_{xx}^-)+2\lambda( \psi_{x}^+ -\sigma \psi_{x}^-)-{\rm i}\kappa(\psi^+ -\sigma\psi^- -u^*\phi)+(1-{\rm i}H)(|u|^2)_x\psi^+$$
$$+2\sigma(1+{\rm i}H)(|u|^2)_x\psi^--u^*({\rm i}\phi_{xx}+2\lambda\phi_x+2u_x\psi^+)-u_t^*\phi=0. \eqno(A.3)$$
 We use (1.4) for the first three terms in (A.3) to simplify it to
 $${\rm i}(u^*\phi)_{xx}+2\lambda(u^*\phi)_{x}+(1-{\rm i}H)(|u|^2)_x\psi^++\sigma(1-{\rm i}H)(|u|^2)_x\psi^-$$
 $$ -u^*({\rm i}\phi_{xx}+2\lambda\phi_x+2u_x\psi^+)-u_t^*\phi=0. \eqno(A.4)$$
 The derivatives $\phi_x$ and $\phi_{xx}$ in (A.4) can be eliminated with the aid of (1.3). After a few manipulations, (A.4) recasts to
 $$\left\{-u_t^*+{\rm i}u_{xx}^*-u^*(1+{\rm i}H)(|u|^2)_x\right\}\phi=0. $$
 Dividing this expression by $\phi$, we obtain the complex conjugate expression  of equation (1.1). \par
 \bigskip
 \noindent {\bf A.2. Conservation laws}\par
 \bigskip
 \noindent We consider the function $u$ analytic in the upper-half complex plane and vanishes rapidly at infinity. To begin with, introduce the eigenfunctions of the
 linear system (1.3)-(1.6) subjected to the boundary conditions $\phi\rightarrow 0,\ \psi^\pm \rightarrow 1$ as $|x|\rightarrow \infty$. Let them be $\bar \phi$ and $\bar\psi^\pm$, respectively.
 Then, equations (1.3)-(1.5) can be written in the form
 $${\rm i}\bar\phi_x+\lambda \bar\phi+u\bar\psi^+=0, \eqno(A.5)$$
$$\bar\psi^+- \bar\psi^--u^*\bar\phi=0, \eqno(A.6)$$
$${\rm i}\bar\phi_t-2{\rm i}\lambda \bar\phi_x+\bar\phi_{xx}-2{\rm i}u_x\bar\psi^+=0, \eqno(A.7)$$
 where we have put $\sigma=1$ and $\kappa=0$ as is consistent with the boundary conditions.
 Applying the projection operator $P_+= {1\over 2}(1-{\rm i}H)$ to 
 (A.6) and using the boundary condition for $\bar\psi^+$ as well as the relations $P_+(\bar\psi^+-1)=\bar\psi^+-1, P_+(\bar\psi^--1)=0$, we deduce  $\bar\psi^+=1+P_+(u^*\bar\phi)$.
   Substituting this expression into (A.5) and (A.7), respectively, we rewrite them as
 $${\rm i}\bar\phi_x+\lambda \bar\phi+uP_+(u^*\bar\phi)+u=0, \eqno(A.8)$$
  $${\rm i}\bar\phi_t-2{\rm i}\lambda \bar\phi_x+\bar\phi_{xx}-2{\rm i}u_xP_+(u^*\bar\phi)- 2{\rm i}u_x=0. \eqno(A.9)$$
  We differentiate (A.8) by $x$ and use it to replace the second term of (A.9). Consequently,  equation (A.9) becomes
  $${\rm i}\bar\phi_t-\bar\phi_{xx}+2{\rm i}uP_+(u^*\bar\phi)_x=0. \eqno(A.10)$$
  It follows from (A.10) and equation (1.1) that
  $${\rm i}(u^*\bar \phi)_t=(u^*\bar\phi_x-u_x^*\bar\phi)_x-{\rm i}(u^*|u|^2\bar\phi)_x+H(|u|^2)_xu^*\bar\phi-|u|^2H(u^*\bar\phi)_x. \eqno(A11)$$
  \par
  Integrating (A.11) with respect to $x$ and taking into account the formula $\int_{-\infty}^\infty\, fHg_x\,dx=\int_{-\infty}^\infty\, gHf_x\,dx$, we find that
  the quantity $I\equiv \int_{-\infty}^\infty u^*\bar \phi\,dx$ is conserved in time.  This can be interpreted as  a generating function for the conserved quantities. Specifically, if we
  expand $\bar\phi$ in inverse powers of $\lambda$ as $\bar\phi=\sum_{j=1}^\infty(-1)^j\bar\phi_j\lambda^{-j}$,  insert this series into (A.8) and then set the coefficients
  of $\lambda^{-j}$ to zero, we obtain the linear recursion relation for $\bar\phi_j$
  $$\bar\phi_1=u, \quad \bar\phi_{j+1}={\rm i}\bar\phi_{j,x}+uP_+(u^*\bar\phi_j), \quad (j=1, 2, ...). \eqno(A.12)$$
  The $j$th conserved quantity $I_j$ is then generated via the series expansion
  $$I=\sum_{j=1}^\infty(-1)^jI_j\lambda^{-j}, \quad I_j=\int_{-\infty}^\infty u^*\bar \phi_j\,dx. \eqno(A.13)$$
  The explicit forms of $\bar\phi_2$ and $\bar\phi_3$ are found to be as
  $$\bar\phi_2={\rm i}u_x+uP_+(|u|^2), \eqno(A.14)$$
  $$\bar\phi_3=-u_{xx}+{\rm i}(uP_+|u|^2)_x+u\left[{\rm i}P_+(u^*u_x)+P_+\{|u|^2P_+(|u|^2)\}\right]. \eqno(A.15)$$
  The  conserved quantities $I_j$  for $j=1, 2, 3$ are computed by using  (A.13). They read
  $$I_1=\int_{-\infty}^\infty |u|^2\,dx, \eqno(A.16)$$
  $$I_2=\int_{-\infty}^\infty \left[{1\over 2}|u|^4+{{\rm i}\over 2}(u^*u_x-u_x^*u)\right]\,dx, \eqno(A.17)$$
  $$I_3=\int_{-\infty}^\infty \left[{1\over 3}|u|^6+{{\rm i}\over 2}(u^*u_x-u_x^*u)|u|^2+{1\over 2}|u|^2H(|u|^2)_x+u_x^*u_x\right]\,dx. \eqno(A.18)$$
  \par
   \bigskip
\noindent{\bf A.3. An equivalent form of the Lax pair} \par
\bigskip
\noindent In [10], the Lax formulation of (1.1) is given in the form
$${\partial L_u\over \partial t}=[B_u, L_u], \eqno(A.19)$$
where the linear operators $L_u$ and $B_u$ acting on the function $\phi$ analytic in the upper-half complex $x$ plane are defined by
$$L_u\phi=-{\rm i}\phi_x-T_uT_{u^*}\phi, \eqno(A.20)$$
$$B_u\phi=-T_uT_{u_x^*}\phi+T_{u_x}T_{u^*}\phi-{\rm i}(T_uT_{u^*})^2\phi, \eqno(A.21)$$
with the Toeplitz operator $T_u\phi=P_+(u\phi)\equiv {1\over 2}(1-{\rm i}H)(u\phi)$. The Lax equation (A.19) stems from the compatibility condition of
the  system of linear PDEs for $\phi$
$$L_u\phi=\lambda\phi, \quad \phi_t=B_u\phi. \eqno(A.22)$$
On the other hand, the linear system (1.1)-(1.6) can be put into a form similar to (A.8) and (A.9). This leads to an alternative form of the Lax equation
$${\partial L\over \partial t}=[B, L], \eqno(A.23)$$
where
$$L\phi\equiv -{\rm i}\phi_x-uP_+(u^*\phi)=\lambda\phi, \eqno(A.24)$$
$$\phi_t=B\phi\equiv {\rm i}\phi_{xx}+2\lambda\phi_x+2u_xP_+(u^*\phi)-{\rm i}\kappa \phi. \eqno(A.25)$$
\par
Now, we show that (A.19) is equivalent to (A.23). First, by invoking the property of the operator $P_+$, we obtain the relations
$$T_uT_{u^*}\phi=uP_+(u^*\phi), \quad T_uT_{u_x^*}\phi=uP_+(u_x^*\phi), \quad T_{u_x}T_{u^*}\phi=u_xP_+(u^*\phi), $$
$$ (T_uT_{u^*})^2\phi=uP_+\{u^*uP_+(u^*u\phi)\}. \eqno(A.26)$$
It immediately follows from (A.20), (A.24) and (A.26) that $L=L_u$. A straightforward computation using (A.24) gives
$$\phi_{xx}={\rm i}u_xP_+(u^*\phi)+{\rm i}uP_+(u_x^*\phi)-uP_+\{u^*uP_+(u^*u\phi)\}-2\lambda uP_+(u\phi)-\lambda^2\phi. \eqno(A.27)$$
Next, substituting (A.24) and (A.27) into (A.25), we deduce
$$B\phi=u_xP_+(u^*\phi)-uP_+(u_x^*\phi)-{\rm i}uP_+\{u^*uP_+(u^*u\phi)\}+{\rm i}\lambda^2\phi-{\rm i}\kappa\phi. \eqno(A.28)$$
Referring to (A.26), (A.21) recasts to
$$B_u\phi=u_xP_+(u^*\phi)-uP_+(u_x^*\phi)-{\rm i}uP_+\{u^*uP_+(u^*u\phi)\}. \eqno(A.29)$$
Last, if we put $\kappa=\lambda^2$ in (A.28), then we find $B=B_u$, completing the proof. \par
In view of the result $L=L_u$, the $j$th conserved quantity from (A.13) is expressed as
$$I_j=(-1)^{j-1}\int_{-\infty}^\infty u^*L^{j-1}udx=(-1)^{j-1}\int_{-\infty}^\infty u^*L_u^{j-1}udx. \eqno(A.30)$$
This reproduces the conserved quantities derived in [11].
 \par
 \bigskip
 \noindent {\bf Appendix B. Proof of lemma 2.1.} \par
  \bigskip
  \noindent First, we enumerate the basic formulas of  determinants which are used frequently in the   proof.   Among them, Jacobi's identity
  will play an important role. See, for example [36]. \par 
  $${\partial |D|\over \partial x}=\sum_{j,k=1}^N{\partial d_{jk}\over \partial x}\,D_{jk}, \eqno(B.1)$$
$$\begin{vmatrix} D & {\bf a}^T\\ {\bf b} & z\end{vmatrix}=|D|z-\sum_{j,k=1}^ND_{jk}a_jb_k,  \eqno(B.2)$$
$$|D({\bf a}, {\bf b}; {\bf c}, {\bf d})||D|= |D({\bf a}; {\bf c})||D({\bf b}; {\bf d})|-|D({\bf a}; {\bf d})||D({\bf b}; {\bf c})|,\ ({\rm Jacobi's\ identity}), \eqno(B.3)$$
$$\delta_{jk}|D|=\sum_{l=1}^Nd_{jl}D_{kl}=\sum_{l=1}^Nd_{lj}D_{lk}, \eqno(B.4)$$
\begin{align}
D_{jk} &= \sum_{l=1}^Nd_{lm}\,D_{jl, km}\quad (k\not=m) \tag{B.5a} \\
&= \sum_{m=1}^Nd_{lm}\,D_{jl, km}\quad (j\not=l), \tag{B.5b}
\end{align}
 \bigskip
  \noindent{\bf B.1. Proof of (2.8a)}\par
  \medskip
  \noindent  We differentiate the element $f_{jk}$ of the matrix $F$ from (2.2a) to obtain
  \begin{align}
  (f_{jk})_t &= {\rm i}(p_j^2-q_j^2)\zeta_j\delta_{jk}  \notag \\
             &= {\rm i}(p_j^2-q_k^2)f_{jk}-{\rm i}(p_j+q_k). \notag
  \end{align}
  Referring to (B.1) and (B.4), we deduce
  \begin{align}
  |F|_t &={\rm i}\sum_{j,k=1}^N(p_j^2-q_k^2)f_{jk}F_{jk}-{\rm i}\sum_{j,k=1}^N(p_j+q_k)F_{jk} \notag \\
        &={\rm i}\sum_{j=1}^N(p_j^2-q_j^2)|F|+{\rm i}(|F({\bf 1}; {\bf p})|+|F({\bf q}; {\bf 1})|). \notag
  \end{align}
\par
\bigskip
 \noindent{\bf B.2.  Proof of (2.8b)}\par
 \medskip
 \noindent   The proof is performed in parallel with the proof of (2.8a). Specifically,
 \begin{align}
  (f_{jk})_x &=- {\rm i}(p_j-q_j)\zeta_j\delta_{jk}  \notag \\
             &= -{\rm i}(p_j-q_k)f_{jk}+{\rm i}. \notag
  \end{align}
\begin{align}
  |F|_x &=-{\rm i}\sum_{j,k=1}^N(p_j-q_k)f_{jk}F_{jk}+{\rm i}\sum_{j,k=1}^NF_{jk} \notag \\
        &=-{\rm i}\mu|F|-{\rm i}|F({\bf 1}; {\bf 1})|. \notag
\end{align}
\par
\medskip
 \noindent{\bf B.3. Proof of (2.8c)}\par
 \medskip
 \noindent It follows by differentiating (2.8b) with respect to $x$ that
$$|F|_{xx}=-{\rm i}\mu|F|_x+{\rm i}\sum_{j,k=1}^N(F_{jk})_x. $$
Applying (B.1) and (B.5) to  the  cofactor $F_{jk}$, we obtain
\begin{align}
 (F_{jk})_x &= \sum_{l,m=1}^N(f_{lm})_xF_{lj,mk} \notag \\
           &=-{\rm i}\sum_{l,m=1}^N(p_l-q_m)f_{lm}F_{lj,mk}+{\rm i}\sum_{l,m=1}^NF_{lj,mk}. \notag
\end{align}
Since $|F|\not=0$, Jacobi's formula (B.3) can be used to give
$$F_{lj,mk}={1\over |F|}(F_{lm}F_{jk}-F_{lk}F_{jm}).$$
If we substitute this into the expression of $(F_{jk})_x $ and use (B.4), we obtain
\begin{align}
(F_{jk})_x &=-{{\rm i}\over |F|}\sum_{l,m=1}^N(p_l-q_m)f_{lm}(F_{lm}F_{jk}-F_{lk}F_{jm})+{{\rm i}\over |F|}\sum_{l,m=1}^N(F_{lm}F_{jk}-F_{lk}F_{jm}) \notag \\
           &=-{\rm i}\mu F_{jk}+{\rm i}(p_j-q_k)F_{jk}+{{\rm i}\over |F|}\sum_{l,m=1}^N(F_{lm}F_{jk}-F_{lk}F_{jm}). \notag 
\end{align}
Thus, we find, after introducing the above expression into the right-hand side of $|F|_{xx}$ and noting the identity, 
$\sum_{j,k,l,m=1}^N(F_{lm}F_{jk}-F_{lk}F_{jm})=0$ 
that
$$|F|_{xx}=-{\rm i}\mu |F|_x+\mu\sum_{j,k=1}^NF_{jk}-\sum_{j,k=1}^N(p_j-q_k)F_{jk}. $$
Last, formula (2.8c) follows by substituting $|F|_x$ from (2.8b) and referring to  (B.2). \par
\bigskip
 \noindent{\bf B.4. Proof of (2.9a)}\par
 \medskip
 \noindent It follows by applying (B.1) and (B.4) that
\begin{align}
|G|_t &= \sum_{j,k=1}^N(g_{jk})_tG_{jk} \notag \\
      &= {\rm i}\sum_{j=2, k=1}^N(p_j^2-q_k^2)g_{jk}G_{jk}-{\rm i}\sum_{j=2, k=1}^N(p_j+q_k)G_{jk} \notag \\
      &={\rm i}\sum_{j=1}^N(p_j^2-q_j^2)|G|+{\rm i}\sum_{k=1}^Nq_k^2g_{1k}G_{1k}- {\rm i}\sum_{j=2, k=1}^N(p_j+q_k)G_{jk}. \notag
 \end{align}
Applying the cofactor expansion to the second term with $g_{1k}=1\ (k=1, 2, ..., N)$ and taking into account the definitions (1.7) and (1.8), we obtain
$$\sum_{k=1}^Nq_k^2q_{1k}G_{1k}=-|F({\bf Q}; {\bf e}_1)|. $$
We compute the third term  while taking into account the relation $p_1=0$ 
\begin{align}
\sum_{j=2, k=1}^N(p_j+q_k)G_{jk}
&=\sum_{j=1, k=1}^N(p_j+q_k)G_{jk}+\sum_{k=1}^Nq_kG_{1k}  \notag \\
&= -|G({\bf 1};{\bf p})|-|G({\bf q};{\bf 1})|+|G({\bf q};{\bf e}_1)|. \notag
\end{align}
Since $p_1=0$, the first term in the above expression becomes zero. The sum of the second and third terms is shown to be $|F({\bf q}, {\bf 1}; {\bf 1}, {\bf e}_1)|$.
Summing up these results, we confirm (2.9a). 
 \par
\bigskip
\noindent{\bf B.5. Proof of (2.9b)}\par
\medskip
\noindent In view of the relations $p_1=0$ and $g_{1k}=1 \ (k=1, 2, ..., N)$, we deduce
\begin{align}
|G|_x &=-{\rm i}\sum_{j=2,k=1}^N(p_j-q_k)g_{jk}G_{jk}+{\rm i}\sum_{j=2,k=1}^NG_{jk} \notag \\
      &=-{\rm i}\mu|G|-{\rm i}\sum_{k=1}^Nq_kG_{1k}+{\rm i}\sum_{j,k=1}^NG_{jk}-{\rm i}\sum_{k=1}^NG_{1k}. \notag
\end{align}
The sum of the third and fourth terms is  found to be zero whereas
the second term can be written as ${\rm i}|F({\bf q};{\bf e}_1)|$, which, plugged into the first term, leads to (2.9b). \par
\bigskip
\noindent{\bf B.6. Proof of (2.9c)}\par
\medskip
\noindent Define an $N\times N$ matrix $\hat G$ by
$$\hat G=(\hat g_{jk})_{1\leq j,k\leq N}, \quad \hat g_{1k}=q_k, \quad \hat g_{jk}=f_{jk},\quad (j=2, 3, ..., N, k=1, 2, ..., N). $$
Then,  $|\hat G|=-|F({\bf q}; {\bf e}_1)|$.
We compute $|\hat G|_x$ by using the rule (B.1) and obtain
\begin{align}
|\hat G|_x &=\sum_{j,k=1}^N(\hat g_{jk})_x\hat G_{jk} \notag \\
           &=-{\rm i}\sum_{j=2,k=1}^N(p_j-q_k)\hat g_{jk}\hat G_{jk}+{\rm i}\sum_{j=2,k=1}^N\hat G_{jk}. \notag 
\end{align}
Referring to  (B.4), we can see that
$$\sum_{j=2,k=1}^Nq_k\hat g_{jk}\hat G_{jk}=\sum_{k=1}^Nq_k|\hat G|-\sum_{k=1}^Nq_k^2\hat G_{1k},$$
which, inserted into the first term of $|\hat G|_x$, gives
\begin{align}
|\hat G|_x &=-{\rm i}\left\{\sum_{j=1}^N(p_j-q_j)|\hat G|+\sum_{k=1}^Nq_k^2\hat G_{1k}\right\} 
             +{\rm i}\sum_{j=2,k=1}^N\hat G_{jk} \notag \\
           &=-{\rm i}\mu|\hat G|+{\rm i}|F({\bf Q};{\bf e}_1)|+{\rm i}(-|\hat G({\bf 1}; {\bf 1})|+|G|), \notag
\end{align}
where we have applied (B.2) in passing to the second line.  It is now straightforward to show that
$$-|\hat G({\bf 1};{\bf 1})|+|G|=-|F({\bf q}, {\bf 1}; {\bf q}, {\bf e}_1)|.$$
If we substitute the above  formulas and $|G|_x$ from (2.9b) into the expression $|G|_{xx}=-{\rm i}(\mu|G|_x+|\hat G|_x)$, we finally arrive at (2.9c). \par
\bigskip
\noindent {\bf Appendix C. Proof of lemma 2.2.}\par
\bigskip
\noindent {\bf C.1. Proof of (2.11)}\par
\bigskip
\noindent Introduce the Cauchy matrix $C$ and its inverse $\tilde C(=C^{-1})$
$$C=(c_{jk})_{1\leq j,k\leq N}, \quad \tilde C=(\tilde c_{jk})_{1\leq j,k\leq N}, \eqno(C.1)$$
where
$$ c_{jk}={1\over p_j-q_k},\quad \tilde c_{jk}=(p_k-q_j)A_k(q_j)B_j(p_k),$$
with
$$A_k(q_j)= \prod_{\substack{l=1\\ (l\not=k)}}^N{q_j-p_l\over p_k-p_l},
\quad B_j(p_k)= \prod_{\substack{l=1\\ (l\not=j)}}^N{p_k-q_l\over q_j-q_l},\quad (j, k=1, 2, ..., N).$$
The following formulas are well-known [37]
$$\sum_{j=1}^N\tilde c_{jk}={\prod_{l=1}^N(p_k-q_l)\over \prod_{\substack{l=1\\ (l\not=k)}}^N(p_k-p_l)}=(p_k-q_k)\tilde c_k,
\quad \tilde c_k=\prod_{\substack{l=1\\ (l\not=k)}}^N{p_k-q_l\over p_k-p_l},\quad (k=1, 2, ..., N), \eqno(C.2)$$
$$|\tilde C|=\prod_{j=1}^N(p_j-q_j)\prod_{\substack{j,k=1\\ (j>k)}}^N{(p_j-q_k)(q_j-p_k)\over (p_j-p_k)(q_j-q_k)}
=\prod_{j=1}^N(p_j-q_j)\prod_{\substack{j,k=1\\ (j>k)}}^Ne^{-A_{jk}}. \eqno(C.3)$$
\par
Now, we are going to  start the proof. First, we compute
\begin{align}
(F^*\tilde C)_{jk} &=\sum_{l=1}^N\left(\zeta_j^*\delta_{jl}+{1\over p_j-q_l}\right)\tilde c_{lk}\notag \\
                   &=\zeta_j^*\tilde c_{jk}+\delta_{jk}. \notag
 \end{align}
and  evaluate the determinant of the matrix $F^*\tilde C$ 
$$|F^*\tilde C|=\left|\left(\zeta_j^*\tilde c_{jk}+\delta_{jk}\right)_{1\leq j,k\leq N}\right|. $$
We substitute the relation 
$$\zeta_j^*={e^{{\rm i}\theta_j+\delta_j}\over p_j-q_j}={e^{2(\delta_j-\phi_j)}\over (p_j-q_j)^2e^{-2\phi_j}\zeta_j},$$
into the above expression and then extract the factor $e^{2\phi_j}/\zeta_j$ from the $j$th row for $j=1, 2, ..., N$ and obtain
$$|F^*\tilde C|=\prod_{j=1}^N{e^{2\phi_j}\over \zeta_j}\left|\left(e^{2(\delta_j-\phi_j)}{\tilde c_{jk}\over (p_j-q_j)^2}+e^{-2\phi_j}\zeta_j\delta_{jk}\right)_{1\leq j,k\leq N}\right|. $$
To modify this further, we insert the relations
$$e^{2(\delta_j-\phi_j)}=\prod_{\substack{l=1\\ (l\not=j)}}^N{(p_j-p_l)(q_j-q_l)\over (p_j-q_l)(q_j-p_l)}, 
\quad \exp\,\left[-\sum_{j=1}^N({\rm i}\theta_j+\phi_j)\right]=\prod_{j=1}^N{(p_j-q_j)\zeta_j\over e^{\delta_j+\phi_j}}, $$
as well as the expression of $\tilde c_{jk}$ from (C.1). This gives
$$|F^*\tilde C|\exp\,\left[-\sum_{j=1}^N({\rm i}\theta_j+\phi_j)\right] $$
\begin{align}
&= \prod_{j=1}^N\left\{(p_j-q_j)e^{\phi_j-\delta_j}\right\}\left|\left({p_k-q_k\over (p_j-q_j)(p_k-q_j)}\prod_{\substack{l=1\\ (l\not=j)}}^N{p_j-p_l\over p_j-q_l}\prod_{\substack{l=1\\ (l\not=k)}}^N{p_k-q_l\over p_k-p_l}    
+e^{-2\phi_j}\zeta_j\delta_{jk}\right)_{1\leq j,k\leq N}\right| \notag \\
&=\prod_{_j=1}^N\left\{(p_j-q_j)e^{\phi_j-\delta_j}\right\}\left|\left(e^{-2\phi_j}\zeta_j\delta_{jk}+{1\over p_j-q_k}\right)_{1\leq j,k\leq N}\right|, \tag{C.4}
\end{align}
where in passing to the second line, we have extracted the factor $(p_j-q_j)^{-1}\prod_{\substack{l=1\\ (l\not=j)}}^N(p_j-p_l)/( p_j-q_l)$ from the $j$th row 
and the factor $(p_k-q_k)\prod_{\substack{l=1\\ (l\not=k)}}^N(p_k-q_l)/( p_k-p_l)$
from the $k$th column, respectively for $j, k=1, 2, ..., N$ and used the formula $|\bar F^T|=|\bar F|$. Last, dividing both sides of (C.4) by $|\tilde C|$ and taking into account  the relations (C.3) and
$ \prod_{j=1}^Ne^{\phi_j-\delta_j}=\prod_{\substack{j,k=1\\ (j>k)}}^Ne^{-A_{jk}}$,
we establish (2.11).
\par
\noindent {\bf C.2. Proof of (2.12)}\par
\medskip
\noindent We  compute the product of the two determinants $|G|^*$ and $|\tilde C|$ and express it in the form of a bordered determinant
$$|G|^*|\tilde C|=-\begin{vmatrix} b_{11} & b_{12}  &\cdots &b_{1N} & 1 \\
                                  b_{21} & b_{22}  &\cdots &b_{2N} & 0 \\
                                  \vdots & \vdots  & \ddots &\vdots & \vdots  \\
                                  b_{N1} & b_{N2} & \cdots & b_{NN} & 0 \\
                                  \sum_{j=1}^N\tilde c_{j1} &\sum_{j=1}^N\tilde c_{j2} & \cdots & \sum_{j=1}^N\tilde c_{jN} &0
                                  \end{vmatrix}, $$
where $B=F^*\tilde C=(b_{jk})_{1\leq j,k\leq N}$ is an $N\times N$ matrix with elements
$$b_{jk}={1\over p_j-q_j}{e^{2\phi_j}\over \zeta_j}{\tilde c_k\over \tilde c_j}{p_k-q_k\over p_k-q_j}+\delta_{jk}.$$
After rewriting the $(N+1)$th row by the formula (C.2), we extract the factor $e^{2\phi_j}/\{(p_j-q_j)\zeta_j\tilde c_j\}$ from the $j$th row
and the factor $(p_k-q_k)\tilde c_k$ from the $k$th column, respectively for $j=2, 3, ..., N$ and $k=1, 2, ..., N$.
The resulting expression can be written as
$$|G|^*|\tilde C|=-\prod_{j=2}^N{e^{2\phi_j}\over \zeta_j}{1\over \tilde c_j}{1\over p_j-q_j}\prod_{k=1}^N\tilde c_k(p_k-q_k)$$
 $$\times\begin{vmatrix} 
(p_1-q_1)^{-1}\tilde c_1^{-1}b_{11}  & (p_2-q_2)^{-1}\tilde c_2^{-1}b_{12} &\cdots 
&(p_N-q_N)^{-1}\tilde c_N^{-1}b_{1N} & 1 \\
                                  \bar f_{12} & \bar f_{22}  &\cdots &\bar f_{N2} & 0 \\
                                  \vdots & \vdots  & \ddots &\vdots & \vdots  \\
                                 \bar f_{1N} &\bar f_{2N}  & \cdots & \bar f_{NN} & 0 \\
                                  1& 1 & \cdots & 1 & 0
                                  \end{vmatrix}, $$
  where the elements $\bar f_{jk}$ have been  defined by (2.10).                                
Interchanging the first row and $(N+1)$th row and then expanding the determinant with respect to $(N+1)$th column, we deduce
$$|G|^*|\tilde C|=(p_1-q_1)\tilde c_1\prod_{j=2}^N{e^{2\phi_j}\over \zeta_j}
 \begin{vmatrix} 1  &1   &\cdots &1 \\
                  \bar f_{12} & \bar f_{22}  &\cdots &\bar f_{N2}  \\
                  \vdots & \vdots  & \ddots &\vdots   \\
                  \bar f_{1N} & \bar f_{2N}  & \cdots & \bar f_{NN}  
 \end{vmatrix},$$
Since $|\tilde C|\not=0$ by (C.3) and (2.4), we can divide the above expression by $|\tilde C|$. If we insert the relations
$$\phi_j=0, \quad \zeta_j={1\over p_j-q_j}\,\exp\,\left[-{\rm i}\theta_j+{1\over 2}\sum_{\substack{k=1\\ (k\not=j)}}^NA_{jk}\right], \quad (j=2, 3, ..., N), \quad p_1=0,
\quad \tilde c_1=\prod_{j=2}^N{q_j\over p_j}, $$
into the result and take into account the relations $\bar f_{jk}=f_{jk}\ (j=1, 2, ..., N, k=2, 3, ..., N)$, we find
$$|G|^*= \prod_{j=2}^N{q_j\over p_j}\,\exp\left[{\rm i}\sum_{j=2}^N\theta_j+{1\over 2}\sum_{k=2}^NA_{1k}\right]d, \eqno(C.5)$$ 
 where a factor  $d$ is the determinant of a matrix constructed from $F^T$ by replacing its first row by the row vector ${\bf 1}$. Explicitly,
$$d=\begin{vmatrix} 1  &1   &\cdots &1 \\
                 f_{12} &  f_{22}  &\cdots & f_{N2}  \\
                  \vdots & \vdots  & \ddots &\vdots   \\
                   f_{1N} &  f_{2N}  & \cdots &  f_{NN}  
 \end{vmatrix}.$$
After a few manipulations,  it can be recast to
$$d =
\begin{vmatrix} f_{22} &\cdots &f_{2N}&1 \\
                \vdots &\ddots &\vdots & \vdots\\
                f_{N2} &\cdots & f_{NN} & 1 \\
                -{1\over q_2}& \cdots &-{1\over q_N}& 1
                \end{vmatrix}, 
                $$
where we have used $f_{1j}=-1/q_j\ (j=2, 3, ... ,N)$.
Then, after  extracting the factor  $1/q_{k+1}$ from the $k$th column
for $k=1, 2, ..., N-1$, we add the $N$th column to the $j$th column for $j=1, 2, N-1$, giving
\begin{align}
d &=\prod_{j=2}^N{p_j\over q_j}\left|\left({q_j\over p_j}\,\zeta_j\delta_{jk}+{1\over p_j-q_k}\right)_{2\leq j,k\leq N}\right| \notag \\
  &=\prod_{j=2}^N{p_j\over q_j}\,|\bar G|. \tag{C.6}
\end{align}
Consequently, substitution of (C.6) into (C.5) yields (2.12).
 \par
 \bigskip
\noindent {\bf D. Proof of proposition 2.4}\par
\medskip
 \noindent  Let  $L_j$ and $l_j$ be
$$L_j=\sum_{k=1}^Nf_{jk}l_k, \quad l_j={\rm i}\phi_{j,x}+\lambda_j\phi_j+u\psi_j^+, \quad (j=1, 2, ..., N). \eqno(D.1)$$
By employing Cramer's rule, the solution of the linear system  (2.32a) can be written as $\phi_j=g_0F_{1j}/|F|$. If we take into account  the definition of the matrix $G$ from (2.2b), we have
 $$\sum_{j=1}^N\phi_j=g_0{|G|\over |F|}=u. \eqno(D.2) $$
Differentiating (2.32a) with respect to $x$ and using the relation 
$$f_{jk,x}=-{\rm i}(p_j-q_k)f_{jk}+{\rm i}, \eqno(D.3) $$
which follows from (2.2a), we deduce
\begin{align}
\sum_{k=1}^Nf_{jk}\phi_x &=-\sum_{k=1}^N f_{jk,x}\phi_k\notag \\
                         &= {\rm i}\sum_{k=1}^N(p_j-q_k)f_{jk}\phi_k-{\rm i}\sum_{k=1}^N\phi_k \notag \\
                         &={\rm i}p_jg_0\delta_{j1}-{\rm i}\sum_{k=1}^Nq_kf_{jk}\phi_k-{\rm i}u \notag \\
                         &=-{\rm i}\sum_{k=1}^Nq_kf_{jk}\phi_k-{\rm i}u, \tag{D.4}
\end{align}
where we have used (D.2) and the relation $p_j\delta_{j1}=p_1\delta_{j1}=0$ with $p_1=0$.  Thus, substitution of (D.4) into (D.1) gives
$$L_j=\sum_{k=1}^Nq_kf_{jk}\phi_k+u+\sum_{k=1}^N\lambda_k f_{jk}\phi_k+u\sum_{k=1}^Nf_{jk}\psi_k^+. $$
Referring to (2.32b) and the relation $\lambda_j=-q_j$, we see that $L_j=0$ for $j=1, 2, ..., N$. 
Since the coefficient matrix $F$ is nonsingular for $x\in\mathbb{R}$\ (See proposition 2.2), this homogeneous  system  for $l_j$ has only the trivial solution $l_j=0 \ (j=1, 2, ..., N)$,
which proves (2.33). \par
Next, define $M_j$ and $m_j$ by
$$M_j=\sum_{k=1}^Nf_{jk}m_k, \quad m_j={\rm i}\phi_{j,t}-2{\rm i}\lambda_j \phi_{j,x}+\phi_{j,xx}-2{\rm i}u_x\psi_j^+, \quad (j=1, 2, ..., N).\eqno(D.5)$$
Use (D.2) and the relation
$$f_{jk,t}={\rm i}(p_j^2-q_k^2)f_{jk}-{\rm i}(p_j+q_k), \eqno(D.6)$$
to verify
\begin{align}
\sum_{k=1}^Nf_{jk}\phi_{k,t} &=-\sum_{k=1}^Nf_{jk,t}\phi_k \notag \\
                             &=-{\rm i}p_j^2g_0\delta_{j1}+{\rm i}\sum_{k=1}^Nq_k^2f_{jk}\phi_k+{\rm i}p_ju+{\rm i}\sum_{k=1}^Nq_k\phi_k. \tag{D.7}
\end{align}
Note that  the first term on the right-hand side of (D.7) vanishes. It now follows from (2.33) with $\lambda_j=-q_j$ that
$$\phi_{k,x}={\rm i}(-q_k\phi_k+u\psi_k^+), \eqno(D.8)$$
$$\phi_{k,xx}=-q_k^2\phi_k+{\rm i}u\psi_{k,x}^++({\rm i}u_x+q_ku)\psi_k^+. \eqno(D.9) $$
By employing (D.8) and (D.9), we compute
$$-2{\rm i}\lambda_k\phi_{k,x}+\phi_{k,xx}-2{\rm i}u_x\psi_k^+=q_k^2\phi_k+{\rm i}u\psi_{k,x}^+-({\rm i}u_x+q_ku)\psi_k^+. \eqno(D.10)$$
Introducing (D.7) and (D.10) into (D.5), $M_j$ becomes
$$M_j=-p_ju-\sum_{k=1}^Nq_k\phi_k+{\rm i}u\sum_{k=1}^Nf_{jk}\psi_{k,x}^++{\rm i}u_x-u\sum_{k=1}^Nf_{jk}q_k\psi_k^+. \eqno(D.11)$$
We differentiate (2.32b) by $x$ and use (D.3) and (2.32b) to obtain
\begin{align}
\sum_{k=1}^N f_{jk}\psi_{k,x}^+ &=-\sum_{k=1}^N f_{jk,x}\psi_k^+ \notag \\
                                &=-{\rm i}p_j-{\rm i}\sum_{k=1}^Nf_{jk}q_k\psi_k^+-{\rm i}\sum_{k=1}^N\psi_k^+. \tag{D.12}
\end{align}
Substituting (D.12) into (D.11), $M_j$ reduces to
\begin{align}
M_j &=-\sum_{k=1}^Nq_k\phi_k+u\sum_{k=1}^N\psi_k^++{\rm i}u_x \notag \\
    &=-{\rm i}\sum_{k=1}^N\phi_{k,x}+{\rm i}u_x, \notag
\end{align}
where in passing to th second line, we have used (D.8). This expression becomes zero by virtue of (D.2), implying that
$l_j=0 \ (j=1, 2, ..., N)$. This proves (2.34). \par
Last, define $N_j$ and $n_j$ by
$$N_j=\sum_{k=1}^Nf_{jk}n_k, \quad n_j={\rm i}\psi^+_{j,t}-2{\rm i}\lambda_j \psi^+_{j,x}+\psi^+_{j,xx}-{\rm i}(1-{\rm i}H)(|u|^2)_x\psi_j^+, \quad (j=1, 2, ..., N).\eqno(D.13)$$
It follows from (2.32b)  that
\begin{align}
\sum_{k=1}^Nf_{jk}\psi_{k,t}^+&= -\sum_{k=1}^Nf_{jk,t}\psi_k^+ \notag \\
                              &= -\sum_{k=1}^N\left\{{\rm i}(p_j^2-q_k^2)f_{jk}-{\rm i}(p_j+q_k)\right\}\psi_k^+ \notag \\
                              &={\rm i}p_j^2+{\rm i}\sum_{k=1}^Nq_k^2f_{jk}\psi_k^++{\rm i}p_j\sum_{k=1}^N\psi_k^++{\rm i}\sum_{k=1}^Nq_k\psi_k^+. \tag{D.14}
\end{align}
If we differentiate (2.32b) twice by $x$  and use (D.3), we can derive the relation
\begin{align}
\sum_{k=1}^Nf_{jk}\psi_{k,xx}^+&=-\sum_{k=1}^Nf_{jk,xx}\psi_k^+-2\sum_{k=1}^Nf_{jk,x}\psi_{k,x}^+ \notag \\
                               &=\sum_{k=1}^N\left\{(p_j-q_k)^2f_{jk}-(p_j-q_k)\right\}\psi_k^+\notag \\
                               &-2{\rm i}\sum_{k=1}^N\left\{-(p_j-q_k)f_{jk}+1\right\}\psi_{k,x}^+. \tag{D.15}
\end{align}
Substituting (D.14) and (D.15) into (D.13) and using (2.32b), we recast  it to
$$N_j=-p_j^2-2p_j\sum_{k=1}^N\psi_k^++\sum_{k=1}^Np_j(p_j-2q_k)f_{jk}\psi_k^+$$
$$-2{\rm i}\sum_{k=1}^N(-p_jf_{jk}+1)\psi_{k,x}^+
+{\rm i}(1-{\rm i}H)(|u|^2)_x. \eqno(D.16)$$ 
With  (D.3), a term in (B.16) is modified as
$$\sum_{k=1}^Nf_{jk}\psi_{k,x}^+=-\sum_{k=1}^N\left\{-{\rm i}(p_j-q_k)f_{jk}+{\rm i}\right\}\psi_k^+. $$
After a few computations, $N_j$ simplifies considerably as shown below:
$$N_j=-2{\rm i}\sum_{k=1}^N\psi_{k,x}^++{\rm i}(1-{\rm i}H)(|u|^2)_x. \eqno(D.17)$$
To proceed, we note the relation 
$$\sum_{k=1}^N\psi_k^+={|F({\bf 1};{\bf 1})|\over |F|}, \eqno(D.18)$$
which can be verified by solving the linear system  (2.32b) by means of Cramer's rule to obtain $\psi_k^+=-\sum_{l=1}^NF_{lk}/|F|$ and then invoking the
formula (B.2).
Furthermore,  the formula
$${\rm i}(1-{\rm i}H)(|u|^2)_x=-2\left({f_x\over f}\right)_x, \eqno(D.19)$$
 follows from (2.5) and an analyticity property of $f(=|F|)$.
 If we substitute (D.18) and (D.19) into (D.17) and use (2.8b), we can see that $N_j=0$ and hence $n_j=0$ for $j=1, 2, ..., N$, which implies (2.35).
Three results established above  now complete the proof of proposition (2.4).
\par
\newpage
 
\leftline{\bf References}\par
\begin{enumerate}[{[1]}]
\item Abanov A G, Bettelheim E and Wiegmann P  2009 Integrable hydrodynamics of Calogero-Sutherland model: bidirectional Benjamin-Ono equation {\it J. Phys. A: Math. Gen.} {\bf 42} 135201
\item Pelinovsky D 1995  Intermediate nonlinear Schr\"odinger equation for internal waves in a fluid of finite depth {\it Phys. Lett.} A {\bf 197} 401-6
\item Matsuno Y 2000 Multiperiodic and multisoliton solutions of a nonlocal nonlinear Schr\"odinger equation for envelope waves {\it Phys. Lett.} A {\bf 278} 53-8
\item Matsuno Y 2001 Linear stability of dark solitary wave solutions of a nonlocal nonlinear Schr\"odinger equation for envelope waves {\it Phys. Lett.} A {\bf 285} 286-92
\item Matsuno Y 2002 Calogero-Moser-Sutherland dynamical systems associated with nonlocal nonlinear Schr\"odinger equation for envelope waves {\it J. Phys. Soc. Japan} {\bf 71} 1415-18
\item Matsuno Y 2003 Asymptotic solutions of the nonlocal nonlinear Schr\"odinger equation in the limit of small dispersion {\it Phys. Lett.} A {\bf 309} 83-9
\item Matsuno Y 2002 Exactly solvable eigenvalue problems for a  nonlocal nonlinear Schr\"odinger equation {\it Inverse Problems} {\bf 18} 1101-25
\item Matsuno Y 2004 A Cauchy problem for the nonlocal nonlinear Schr\"odinger equation {\it Inverse Problems} {\bf 20} 437-45
\item Matsuno Y 2004 New representations of multiperiodic and multisoliton solutions for a class of nonlocal soliton equations {\it J. Phys. Soc. Japan} {\bf 73} 3285-93
\item Tutiya Y 2006 Bright $N$-solitons for the intermediate nonlinear Schr\"odinger equation {\it J. Nonl. Math. Phys.} {\bf 16} 7-23
\item G\'erard P and Lenzmann E 2022 The Calogero-Moser derivative nonlinear Schr\"odinger equation (arXiv: 2208.04105v1[math.AP])
\item Zakharov V E and Shabat A B 1972 Exact theory of two-dimensional self-focusing and one-dimensional self-modulation of waves in nonlinear media {\it Sov. Phys.-JETP} {\bf 34} 62-9
\item Zakharov V E and Shabat A B 1973 Interaction between solitons in a stable medium {\it Sov. Phys.-JETP} {\bf 37} 823-8
\item Fadeev L D and Takhtajan L A 2007 {\it Hamiltonian Methods in the Theory of Solitons}  (Berlin:  Springer)
\item Berntson B K and Fagerlund A 2022 A focusing -defocusing intermediate nonlinear Schr\"odinger system (arXiv: 2212.03751v1[nlin.SI])
\item Matsuno Y 1984 {\it Bilinear Transformation Method} (New York:  Academic)
\item Hirota R 2004 {\it The Direct Method in Soliton Theory} (Cambridge: Cambridge University Press)
\item Dobrokhotov S Yu and Krichever I M  1991 Multiphase solutions of the Benjamin-Ono equation and their averaging {\it Math. Notes} {\bf49} 583-94
\item Dubrovin B A 1975   Periodic problems of the Korteweg-de Vries equation in the class of finite-zone potentials {\it Func. Anal. Appl.} {\bf 9} 215-23
\item Calogero F 1971 Solution of the one-dimensional $N$-body problems with quadratic and/or inversely  quadratic pair potentials {\it J. Math. Phys.} {\bf 12} 419-36
\item Sutherland B 1972 Exact results for a quantum many-body problems in one dimension. II {Phys. Rev.} {\rm A} {\bf 5} 1372-76
\item Moser J 1975 Three integrable Hamiltonian systems connected to isospectral deformations {\it Adv. Math.} {\bf 16} 197-220
\item Olshhanetsky M A and Perelomov A M 1981 Classical integrable finite dimensional system related to Lie algebras {\it Phys. Rep.} {\bf 71} 313-400
\item Polychronakos A P 1995 Waves and solitons in the continuum limit of the Calogero-Sutherland model {\it Phys. Rev. Lett.} {\bf 74} 5153-7
\item van Diejen J F and Vinet L (ed)  2000 {\it Calogero-Moser-Sutherland models (CRM Series in Mathematical Physics)} (New York: Springer)
\item Matsuno Y 1979 Exact multi-soliton solution of the Benjamin-Ono equation {\it J. Phys. A: Math. Gen.} {\bf 12} 619-21
\item Matsuno Y 1980 Interaction of the Benjamin-Ono solitons {\it J. Phys. A: Math. Gen.} {\bf 13} 1519-36
\item Matsuno Y 1986 $N$-soliton solution for the sine-Hilbert equation {\it Phys. Lett.} A {\bf 119} 229-33
\item Matsuno Y 1987 Periodic problem for the sine-Hilbert equation {\it Phys. Lett.} A {\bf 120} 187-90
\item Matsuno Y 1987 Kinks of the sine-Hilbert equation and their dynamical motions {\it J. Phys. A: Math. Gen.} {\bf 20} 3587-606
\item Santini P M, Ablowitz M J and  Fokas A S 1987 On the initial value problem for a class of nonlinear integral evolution equations including the sine-Hilbert equation {\it J. Math. Phys.} {\bf 28} 2310-16 
\item Santini P M 1993  Integrable singular integral evolution equations {\it Important Developments in Soliton Theory} ed A S Fokas and V E Zakharov (New York: Springer) 147-77
\item Matsuno Y 1995 Dynamics of interacting algebraic solitons {\it Int. National. J. Mod. Phys.} {\rm B} {\bf 9} 1985-2081
\item Pelinovsky D E and Grimshaw R H J 1995 A spectral transform for the intermediate nonlinear Schr\"odinger equation {\it J. Math. Phys.} {\bf 36} 4203-19
\item Matsuno Y 2001 $N$-soliton formulae for the intermediate nonlinear Schr\"odinger equation {\it Inverse Problems} {\bf 17} 501-14
\item Vein R and Dale P 1999 {\it Determinants and Their Applications in Mathematical Physics} (New York: Springer)
\item Schechter S 1959 On the inversion of certain matrices {\it Mathematical Tables and Other Aids to Computation} {\bf 13} 73-7.

\end{enumerate}
\end{document}